\documentclass[aps,pra,twocolumn, longbibliography]{revtex4-1} 
\usepackage{verbatim}
\usepackage{dsfont} 
\usepackage{amsmath}
\usepackage{amsthm} 
\usepackage{enumerate}
\usepackage{graphicx}
\usepackage{mathtools} 
\usepackage{bm} 
\usepackage{amssymb} 
\usepackage{hyperref} 
\hypersetup{
	colorlinks=true,
	linkcolor=blue,
	citecolor=blue,
	unicode=true,          
}
\usepackage[capitalize,nameinlink]{cleveref}

\usepackage{blindtext}

\theoremstyle{definition}
\newtheorem{result}{Result}

\newtheorem{prop}{Proposition}
\newtheorem{definition}{Definition}

\newtheorem{corollary}{Corollary}

\newtheorem*{result*}{Result}

\DeclareMathOperator{\Tr}{Tr}

\newcommand{\M}{\mathsf{M}}

\newcommand{\state}{\mathcal{S}} 
\usepackage[bbgreekl]{mathbbol}
\DeclareSymbolFontAlphabet{\mathbb}{AMSb}
\DeclareSymbolFontAlphabet{\mathbbl}{bbold}
\DeclareMathSymbol{\bbepsilon}{\mathord}{bbold}{"0F}

\usepackage{mathrsfs}

\newcommand{\ang}[1]{\left\langle #1 \right\rangle}

\newcommand{\mc}[1]{\mathcal{#1}}

\newcommand{\mb}[1]{\mathbb{#1}}

\newcommand{\R}{\mathbb R}

\newcommand{\T}{\mathbb T}

\usepackage{calc} 
\usepackage{accents}

\makeatletter
\newcommand{\oset}[3][0ex]{%
	\mathrel{\mathop{#3}\limits^{
			\vbox to#1{\kern-2\ex@
				\hbox{$\scriptstyle#2$}\vss}}}}
\makeatother

\usepackage{diagbox}
\usepackage{textpos}
\usepackage{pifont}
\usepackage[dvipsnames,table,xcdraw]{xcolor} 
\usepackage{slashbox}
\usepackage[abs]{overpic}

\definecolor{darker}{RGB}{2.0,100.0,20.0}
\definecolor{punainen}{RGB}{225.0,64.0,64.0}

\newcommand{\punainen}[1]{\textcolor{punainen}{#1}}

\newcommand{\EE}{\mathcal{E}}
\newcommand{\NN}{\mathsf N}
\usepackage{enumitem}

\begin{document}
	
\title{Multivariate R\'enyi divergences characterise betting games with multiple lotteries}

\author{Andr\'es F. Ducuara$^{1,2,3,4}$} 
\email[]{andres.dg@ntu.edu.sg}

\author{Erkka Haapasalo$^{5}$}
\email[]{cqteth@nus.edu.sg}

\author{Ryo Takakura$^{6,7}$}
\email[]{takakura.ryo.qiqb@osaka-u.ac.jp}

\affiliation{$^{1}$Nanyang Quantum Hub, School of Physical and Mathematical Sciences, Nanyang Technological University, 637371, Singapore
    \looseness=-1}

\affiliation{$^{2}$Centre for Quantum Technologies, Nanyang Technological University, 637371, Singapore
    \looseness=-1}

\affiliation{$^{3}$Yukawa Institute for Theoretical Physics, Kyoto University, Kitashirakawa Oiwakecho, Sakyo-ku, Kyoto 606-8502, Japan
    \looseness=-1}

\affiliation{$^{4}$Center for Gravitational Physics and Quantum Information, Yukawa Institute for Theoretical Physics, Kyoto University
    \looseness=-1} 

\affiliation{$^{5}$Centre for Quantum Technologies, National University of Singapore, 117543, Singapore 
    \looseness=-1}

\affiliation{$^{6}$Center for Quantum Information and Quantum Biology, The University of Osaka, 1-2 Machikaneyama, Toyonaka, Osaka 560-0043, Japan\looseness=-1}
\affiliation{$^{7}$Graduate School of Science, The University of Osaka, 1-1 Machikaneyama, Toyonaka, Osaka 560-0043, Japan\looseness=-1}

\date{\today}
\begin{abstract}
    We provide an operational interpretation of the multivariate R\'enyi divergence in terms of economic-theoretic tasks based on betting, risk aversion, and multiple lotteries. Specifically, we show that the multivariate R\'enyi divergence
    $
    D_{\underline{\alpha}}
    (\vec{P}_X)
    $
    of probability distributions
    $
    \vec{P}_X 
    =
    ( 
    p^{(0)}_X,\ldots,p^{(d)}_X 
    )
    $
    and real-valued orders $\underline{\alpha} = (\alpha_0, \dots, \alpha_{d})$ quantifies the economic-theoretic value that a rational agent assigns to $d$ lotteries with odds $o^{(k)}_X \propto (p_X^{(k)})^{-1}$ ($k=1, \dots, d$) on a random event described by $p^{(0)}_X$. 
    In particular, when the odds are fair and the rational agent is allowed to maximise over all betting strategies, the economic-theoretic value (precisely the isoelastic certainty equivalent) that the agent assigns to the lotteries is exactly given by
    $
    w^{\rm ICE}_{\underline{R}}
    =
    \exp[
    D_{\underline{\alpha}}
    (\vec{P}_X)
    ]
    $, where
    $\underline{R}=(R_1,\ldots,R_d)$ is a risk-aversion vector with $R_k = 1+\alpha_{k}/\alpha_0$ being
    the risk-aversion parameter associated with lottery $k$. Furthermore, we introduce a new conditional multivariate R\'enyi divergence that characterises a generalised scenario where the rational agent is allowed to have access to side information. 
    We prove that this new quantity satisfies a data processing inequality which can be interpreted as the increment in the economic-theoretic value provided by side information; crucially we show that such a data processing inequality is a consequence of the agent's economic-theoretically consistent risk-averse attitude towards every lottery and vice versa. 
	Moreover, we demonstrate the applicability of these results to the resource theory of informative measurements within the operational framework of general probabilistic theories (GPTs) encompassing quantum theory. 
	By establishing quantitative connections between information theory, physical theories, and theories from economics, our framework provides a novel operational foundation for quantum state betting games with multiple lotteries in the realm of quantum resource theories.
\end{abstract} 
\maketitle
\begin{textblock*}{3cm}(17cm,-14.5cm)
    \footnotesize YITP-25-40
\end{textblock*}
\vspace{-1cm}
\section{Introduction}
	
Expected utility theory (EUT) is a theoretical framework first formalised during the 1940s in the work of von Neumann and Morgenstern within the theory of games and economic behaviour \cite{risk_vNM}. EUT primarily deals with the development of mathematical models describing the behaviour of rational agents when facing decision problems involving uncertainty \cite{risk_EGS}. A central object of study within EUT is the \emph{utility function}, a real-valued function representing the rational agent's level of satisfaction when acquiring ``desirable assets" such as wealth, goods, or services. A highly non-trivial insight first put forward by economists during the 1950s establishes that the economic-theoretic concept of \emph{risk aversion} is mathematically encoded in the curvature of the utility function \cite{risk_bernoulli, risk_arrow, risk_pratt}. This realisation in turn opened the door in particular for the development of risk-aversion measures, thus establishing a concise mathematical framework for the quantification of risk aversion. The exploration of subjects related to risk aversion has remained a topic of active research in the economic sciences, with several of these efforts having been recognised with the Sveriges Riksbank Prize in Economic Sciences in Memory of Alfred Nobel \cite{nobel}.

In addition to encoding the level of risk aversion of a rational agent, the utility function also resolves decision problems that involve \emph{fixed} amounts of wealth, goods, or services. This follows from the fundamental guiding principle in the economic sciences which establishes that rational agents are utility maximisers and, consequently, they will naturally choose the alternative that provides the largest utility \cite{risk_vNM}. A more involved situation arises when the gains in question involve \emph{uncertainty} or, in other words, when rational agents face decision problems with uncertain gains such as lotteries. In this more general regime, the concept known as the \emph{certainty equivalent} is required. The certainty equivalent of a lottery represents the minimum certain amount of wealth that the rational agent is willing to accept in order to be persuaded not to participate in the given lottery. In other words, it represents the minimum \emph{certain} amount of wealth that is \emph{equivalent} (from the viewpoint of the rational agent) to the lottery. The certainty equivalent thus represents the economic-theoretic ``value" that the rational agent assigns to a given lottery and consequently completely resolves decision problems involving uncertainty; the rational agent would choose the lottery that provides the largest certainty equivalent. The economic-theoretic concepts of decision-making rational agents, risk aversion, and the certainty equivalent, are ubiquitous ideas that have emerged and found usefulness in various scientific disciplines including biology and behavioural ecology \cite{risk_biology1, risk_biology2}, neuroscience \cite{NS1, NS2, NS3}, thermodynamics \cite{EUT_thermo0, EUT_thermo1, EUT_thermo2, EUT_thermo3, EUT_thermo4, EUT_thermo5, EUT_thermo6, EUT_thermo7}, and information theory \cite{kelly, BLP2020, risk_soklakov, QBG_2022, Kamatsuka2022,EUT_info1, EUT_info2,Kamatsuka2026}, with the latter being the main subject of study in this work.

Information theory, first formalised during the 1940s in the works of Shannon \cite{shannon1}, can, pragmatically speaking, be described as the sub-area of applied probability theory that deals with the operational and physical interpretation of information-theoretic quantities such as: unconditional and conditional entropies, mutual information measures, capacities, unconditional and conditional divergences, amongst many others \cite{CT, Tomamichel2016Book}. Amongst this zoo of quantities, the quantity known as the R\'enyi divergence stands out as a cornerstone quantity that acts as a ``parent" to many of the above-mentioned quantities, in addition to having a multitude of applications within the information sciences \cite{renyi,van_Erven_et_al_2014}. The R\'enyi divergence $D_\alpha (p_X^{(0)} || p_X^{(1)})$ of order $\alpha \in [0,1) \cup (1,\infty)$ of probability distributions $p_X^{(0)}$ and $p_X^{(1)}$ (on an alphabet $\mc X$) reads
\begin{align*}
    &D_\alpha(
    p_X^{(0)}
    \|
    p_X^{(1)}
    )=
    \frac{1}{\alpha-1}
    \log
    \left[
    \sum_{x\in \mathcal{X}}
    (
    p_X^{(0)}(x)
    )^\alpha 
    (
    p_X^{(1)}(x)
    )^{1-\alpha}
    \right]
\end{align*}
and can roughly be interpreted as the level of dissimilarity between $p_X^{(0)}$ and $p_X^{(1)}$. In particular, the case $\alpha=1$ recovers the Kullback-Leibler divergence (also known as the relative entropy) \cite{KL1951, CT}. The R\'enyi divergence has found extensive use in information science \cite{renyi, Csiszar1995, Harremoes2006}. Similarly, the extension of the R\'enyi divergence to the realm of quantum mechanics \cite{Petz_85, Petz_1986,Hiai_Petz_91, Muller-Lennert_et_al_2013, Jaksic2012, Audenaert_Datta_2015}, has been an active and fruitful area of research \cite{Mosonyi_Hiai_2011, Carlen_et_al_2018, Zhang2020, Tomamichel2016Book}. Recently, a \emph{multivariate} extension of the R\'enyi divergence taking multiple input probability distributions 
$
\vec{P}_X
=
( 
p_X^{(0)},\ldots,p_X^{(d)}
)
$
as well as multiple orders 
$\underline{\alpha} 
=
(\alpha_0, \dots, \alpha_{d})\in\R^{d+1}$ has received significant attention \cite{mosonyi2024geometric, MV2, Furuya_et_al_2023, farooq2024, Verhagen_et_al_2024, MV4, MV5}. This multivariate R\'enyi divergence is given by
\begin{align*}
    D_{\underline{\alpha}}
    (\vec P_{X})=
    \frac{1}{
        \displaystyle
        \max_{0\le k\le d}\alpha_{k}-1
    }
    \log
    \left[        
    \sum_{x\in \mathcal{X}}
    \prod_{k=0}^{d} 
    \big(
    p_{X}^{(k)}(x)
    \big)^{\alpha_k}
    \right]
    .
\end{align*}
The orders $\underline{\alpha}$ are required to satisfy $\sum_{k=0}^{d} \alpha_k =1$ and either (i) $\alpha_k\geq0$ for all $k=0, \ldots, d$, or
(ii) $\alpha_l\!>\!1$ for some $l$ and $\alpha_{k}\!\leq \!0$ for the other $k\!\neq\! l$.
These requirements are essential for $D_{\underline{\alpha}}$ to be an information-theoretically reasonable divergence.  
In particular, with these conditions for $\underline{\alpha}$, the quantity $D_{\underline{\alpha}}(\vec{P}_X)$ satisfies the data processing inequality with respect to postprocessings of the probability distributions $\vec{P}_X$, which is one of the most primitive properties that a divergence should exhibit.
Interestingly, it was observed that under the same restrictions, the converse of the data processing inequality also holds in a certain sense, that is, a set of probability distributions $\vec{P}_X$ can ``simulate'' another set $\vec{Q}_X$ if  $D_{\underline{\alpha}}(\vec{P}_X)\ge D_{\underline{\alpha}}(\vec{Q}_X)$ holds (for the precise statement, see \cite{farooq2024}). The exploration of the multivariate R\'enyi divergence and adjacent topics is an active area of research which is arguably still in its infancy, with various avenues for research to potentially be uncovered and developed.

\begin{figure}[t!]
    \centering
    \includegraphics[scale=0.43]{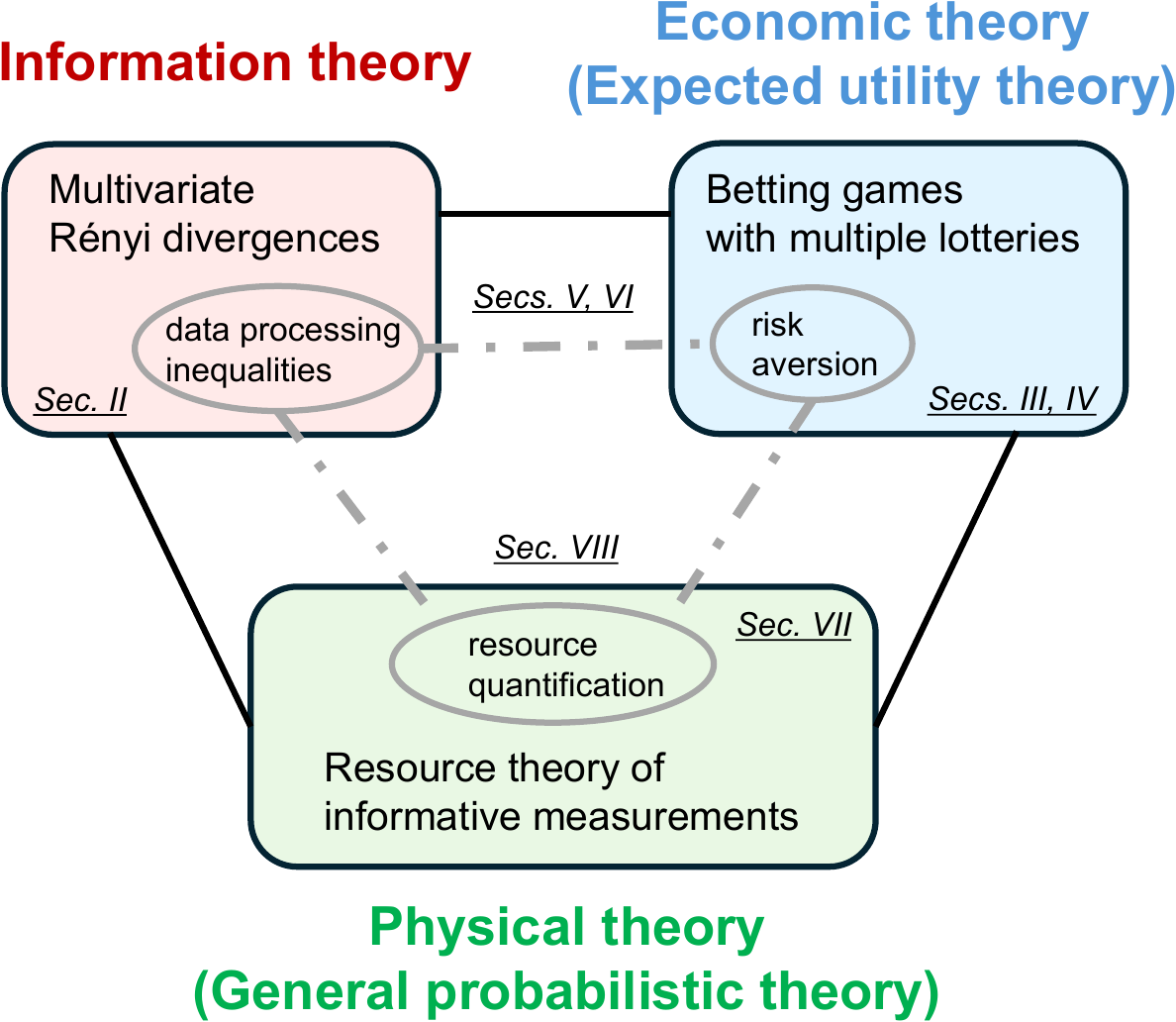}
    \caption{
        Summary of the current work.
        We study three major themes in colored boxes and discuss their quantitative connections represented by gray lines.
        Technical details will be found in the underlined section attached with each topic. 
    }
    \label{fig:summary}
\end{figure}

In this work we show that the multivariate R\'enyi divergence characterises operational tasks involving decision-making rational agents, risk aversion, as well as betting on multiple lotteries. 
The operational task of betting on horse races, or \emph{horse betting games} for short, was first introduced and characterised by Kelly in 1956 within the realm of classical information theory \cite{kelly} (the reader is also referred to \cite{CT,LN_Moser} for a modern treatment).
Using the language of economics and finance, more specifically the theory of expected utility, the original version of these games can be viewed as being played by a \emph{risk-averse} rational agent. 
Recently, Bleuler, Lapidoth, and Pfister \cite{BLP2020} extended the treatment of horse betting games to allow rational agents to display more general attitudes towards risk such as different degrees of \emph{risk aversion}, as well as \emph{risk-neutral}, and \emph{risk-seeking} tendencies. Later in \cite{QBG_2022}, the central role of the ``certainty equivalent" was put forward, and horse betting was further extended to the realm of quantum information theory, taking the form of \emph{quantum state betting} (QSB) games. This development draws inspiration from horse betting as well as standard quantum state discrimination (QSD), the latter being an operational task introduced during the 1970s in the works of Holevo and Helstrom in the context of quantum information theory \cite{HHthm1, HHthm2}. Building on this interdisciplinary line of research, in this work we take one further step and introduce operational tasks involving risk aversion, betting on multiple lotteries, and show that these are characterised by the multivariate R\'enyi divergence, as well as by a new \emph{conditional} multivariate R\'enyi divergence which we propose. 
We furthermore apply these insights to the resource theory of measurement informativeness within \emph{general probabilistic theories}, namely the broadest framework for describing physical phenomena that occur probabilistically. 
We schematically summarise these contributions in \cref{fig:summary}, and we succinctly describe them  in what follows, together with the organisation of this manuscript.


\subsection{Organisation and summary of results}

In \cref{s:s2}, we start by addressing the unconditional multivariate R\'enyi divergence, followed by the introduction of a new \emph{conditional} multivariate R\'enyi divergence inspired by the bivariate Bleuler-Lapidoth-Pfister (BLP) conditional R\'enyi divergence \cite{BLP2020}. As mentioned above, for the (unconditional) multivariate R\'enyi divergence $D_{\underline{\alpha}}$ to be a proper information-theoretic divergence, the orders $\underline{\alpha}$ must satisfy certain conditions. This new conditional quantity is defined for the same admissible orders and, with these orders, we show that it satisfies various data processing inequalities. In \cref{s:s3} we review expected utility theory (EUT), and introduce the concepts of decision problems, risk aversion, the certainty equivalent, and the standard scenario for betting games with a single lottery.

In \cref{s:s4a} we start with our main results by introducing new operational tasks in the form of horse betting games with \emph{multiple} lotteries, where a gambler receives multiple commodities as rewards from bets being placed on the winning horse in a horse race. We address these tasks as betting games with multiple lotteries, or \emph{multi-lottery betting games} for short. In \cref{s:s4r1}, we show that the economic-theoretic value that a rational agent assigns to a multi-lottery betting game can be written in terms of the unconditional multivariate R\'enyi divergence. 
In particular, the admissible order constraints of the multivariate R\'enyi divergence can be reinterpreted as constraints on the gambler's risk-averse attitude towards the lotteries in question. This elucidates a direct link between the information-theoretic consistency of the multivariate divergence and the economic-theoretic consistency of the gambler's attitude towards risk. 
In \cref{s:s4r2}, we take one step forward and consider a generalised scenario where the gambler is allowed to place bets after having access to side information. 
We reveal that the economic-theoretic value that the gambler assigns to the lotteries in this extended scenario can be written in terms of the new conditional multivariate R\'enyi divergence introduced in \cref{s:s2}. 
In particular, we show that the data processing inequality captures the increment in economic-theoretic value due to the availability of side information, thus establishing an explicit quantitative connection between information-theoretic quantities on the one hand, and betting games within the theory of expected utility from the economic sciences on the other.

In \cref{s:gpt1} we further extend these ideas by introducing the operational tasks of state betting games within the framework of \emph{general probabilistic theories} (\emph{GPTs}) \cite{Hardy2001, Barnum2007, Barrett2007, Chiribella2010, Chiribella2011, Masanes2011, Barnum2012}. GPTs provide the most general description for probabilistic events involving states and measurements, including the theory of quantum mechanics as a particular case. A state betting game is constituted by replacing the classical horse race with an ensemble of states of a GPT, and considering the gambler being allowed to perform a measurement to extract side information from the received state. 
We show how the connection between the conditional multivariate Rényi divergence and the economic value of a multi-lottery state betting game can be formulated in terms of states and measurements. In \cref{s:gpt2} we show the applicability of these insights to the resource theory of informative measurements \cite{PS_NL_2019, FB1, FB2, RT_BS_2019}. In particular, we propose a new resource measure for measurement informativeness based on multivariate Rényi divergences, and interpret this measure from an economic-theoretic perspective. These latter results in particular allow for the extension of quantum state betting games to scenarios involving multiple lotteries within the realm of quantum resources theories \cite{Chitambar2019,Gour2025}. 
Overall, we establish explicit connections between information theory, physical theories, and theories of economics by establishing quantitative relations between data processing inequalities, the concept of risk aversion, and resource quantifiers. In \cref{s:conclusions} we conclude with a summary and future perspectives.
	
\section{Multivariate R\'enyi divergences}
\label{s:s2}

In this section we introduce unconditional and conditional multivariate R\'enyi divergences, and derive various data processing inequalities. 

\subsection{Some remarks on notation}

The notation to be used is as follows. 
Consider random variables ($X, G,\ldots$) with finite alphabets (or outcome sets) ($\mathcal{X}, \mathcal{G},\ldots$). 
In this paper, the underlying probability spaces of those random variables are not explicitly shown because they do not affect the argument.
The probability mass function (PMF) $p_X\colon\mc X\to[0,1]$ of a random variable $X$ is defined such that $p_X(x)=\mathrm{Pr}[X=x]$ ($\forall x \in \mathcal{X}$).
We also call a function $p_X\colon \mc X\to[0,1]$ a PMF if it satisfies $p_X(x) > 0$ ($\forall x \in \mathcal{X}$) and $\sum_{x \in \mathcal{X}} p_X(x) = 1$.
In this case, the subscript $X$ is simply an index highlighting that $p_X$ is a function on $\mc X$, and we often do not strictly distinguish these two usages.
Similarly, a function $f\colon\mc X\to\R$ on $\mc X$ is often expressed as $f_X$.
Its expectation value with respect to a PMF $p_X$ is denoted by
$
\mb E_{p_X}[f_X]
\coloneqq
\sum_{x\in \mathcal{X}} 
f_X(x)p_X(x)
$. 
Our analysis in this paper also focuses on $(d+1)$ PMFs on the same alphabet $\mathcal{X}$.
In this case, the $k$th PMF is written as $p_X^{(k)}$ ($k=0,\ldots,d$).
We note that any PMF $p_X$ is assumed to be with full support ($p_X(x) > 0$, $\forall x \in \mathcal{X}$).
In \cref{a:contrem}, we loosen the assumptions that outcome sets are finite and PMFs are with full support, and study more general cases. 
Joint and conditional PMFs are denoted by $p_{XG}$, $p_{G|X}$, respectively. 
When considering multiple conditional PMFs, we use the expression 
$p_{X|G}^{(k)}$ ($k=0,\ldots,d$) similarly to the unconditional case.

For finite sets $\mc X$ and $\mc Y$, a stochastic operator (or Markov kernel) $T_{Y|X}$ can be understood as a set of conditional PMFs $\{t_{X|Y}(\cdot|x)\}_{x\in \mc X}$ on $\mc Y$ or simply a stochastic matrix.
It maps a PMF $p_{X}$ on $\mc X$ to a PMF $q_{Y}$ on $\mc Y$ by
\begin{equation}\label{eq:markov}
    q_{Y}(y)=\sum_{x\in \mc X}t_{Y|X}(y|x)p_{X}(x),
\end{equation}
In this case $q_{Y}$ is written as $q_Y=T_{Y|X}(p_X)$ and said to be obtained through postprocessing $p_{X}$.
We also use the pseudo-inverse $T^\dagger_{X|Y}$ of $T_{Y|X}$, which is a set of conditional PMFs $\{t^\dagger_{X|Y}(\cdot|y)\}_{y\in \mc Y}$ on $\mc X$ defined in terms of \cref{eq:markov} by
\begin{equation*}
    t^{\dagger}_{X|Y}(x|y)=\frac{p_{X}(x)}{q_{Y}(y)}t_{Y|X}(y|x).
\end{equation*}
This pseudo-inverse is used when postprocessing the conditioning systems of conditional PMFs (see \cref{l:l2}).
    
\subsection{Multivariate R\'enyi divergences}

We now introduce an extension of the R\'enyi divergence allowing multiple input PMFs.
\begin{definition}
    [\emph{Unconditional multivariate R\'{e}nyi divergence \cite{farooq2024,Verhagen_et_al_2024}}]
    \label{def:divergences}
    Let $\vec P_{X} = \big( p_X^{(0)},\ldots,p_X^{(d)} \big)$ be $(d+1)$ PMFs with full support.
    The unconditional multivariate R\'{e}nyi divergence of $\vec P_X$ and of orders $\underline{\alpha} = (\alpha_0, \ldots, \alpha_{d}) \in \R^{d+1}$ is denoted as $D_{\underline{\alpha}}(\vec P_{X})$ and given by
    \begin{align*}
        \label{eq:Renyi_mv}
        D_{\underline{\alpha}}
        (\vec P_{X})
        &\coloneqq
        \frac{1}{
            \alpha_{\star}-1
        }
        \log
        \left[        
        \sum_{x\in \mathcal{X}}
        \prod_{k=0}^{d} 
        \big(
        p_{X}^{(k)}(x)
        \big)^{\alpha_k}
        \right]
        ,
    \end{align*}
    where $\alpha_{\star}
    \coloneqq \max_{0\leq k\leq d}
    \alpha_k$. The orders $\underline{\alpha} = (\alpha_0, \ldots, \alpha_{d})$ are real parameters satisfying $\sum_{k=0}^{d} \alpha_k =1$ and either of the following two conditions:
    \begin{equation}
        \label{eq:parameter}
        \begin{aligned}
            \mathrm{(i)} &\ \mbox{$\alpha_k\geq0$ 
                for all 
                $k=0, \ldots, d$},\\
            \mathrm{(ii)} &\ \mbox{$\alpha_l\!>\!1$ for some $l$ and $\alpha_{k}\!\leq \!0$ for the other $k\!\neq\! l$}
            .
        \end{aligned}
    \end{equation}
\end{definition}

\begin{figure}[h!]
    \begin{center}
        \begin{overpic}[scale=0.37, unit=1mm]{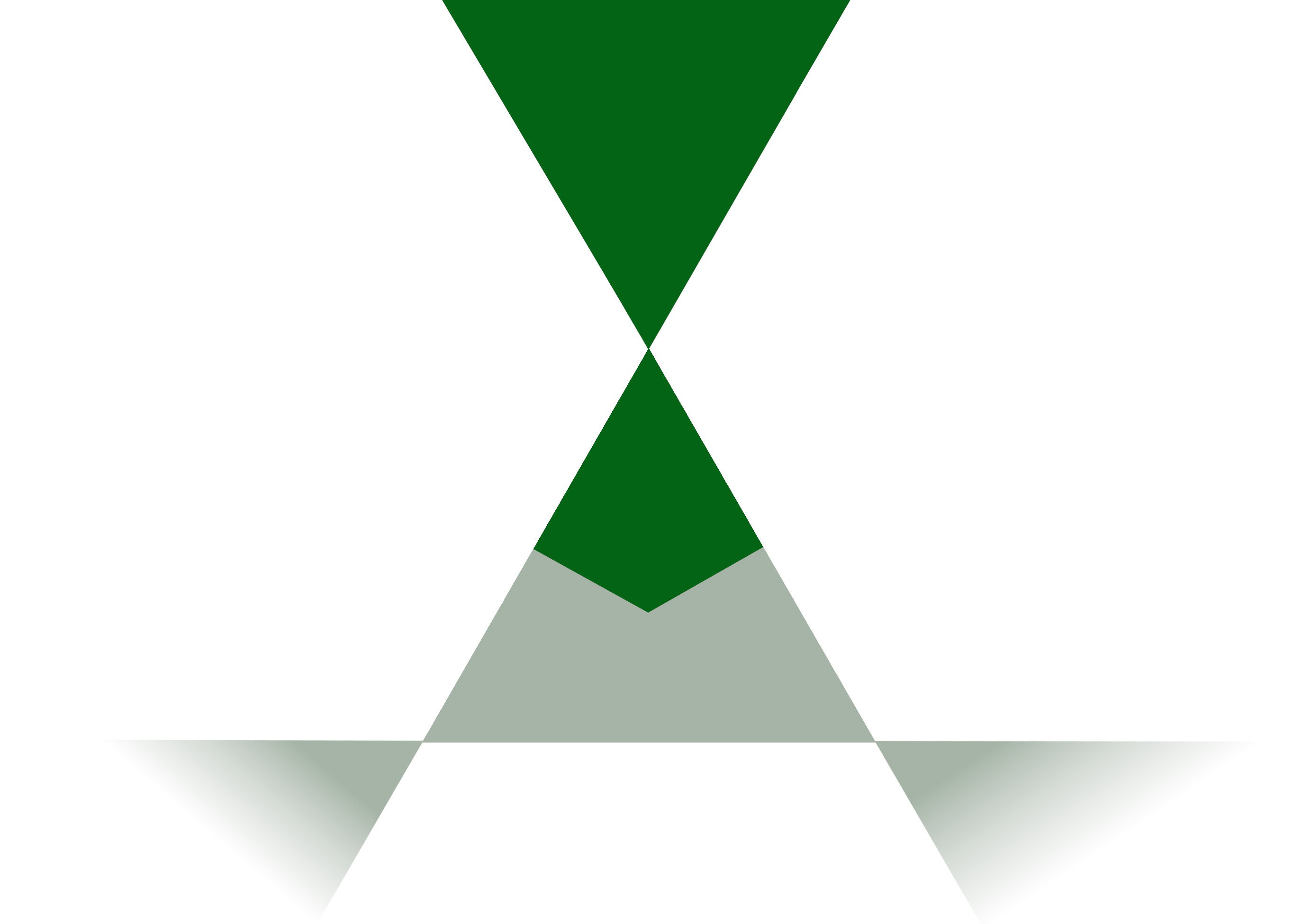}
            \put(43,37){\begin{large}
                    $e_0$
            \end{large}}
            \put(23,13){\begin{large}
                    $e_1$
            \end{large}}
            \put(57,13){\begin{large}
                    $e_2$
            \end{large}}
        \end{overpic}
    \end{center}
    \caption{
        \label{fig:parameterregion} 
        The area of allowed parameter vectors $\underline{\alpha}=(\alpha_0,\alpha_1,\alpha_2)\in\R^3$ for the multivariate R\'enyi divergence $D_{\underline{\alpha}}$ in the case $d=2$. The area is confined onto the affine plane $\sum_{k=0}^2\alpha_k=1$ in $\R^3$. The special points fixing this plane are $e_0=(1,0,0)$, $e_1=(0,1,0)$, and $e_2=(0,0,1)$. 
    }
\end{figure}

In \cref{fig:parameterregion} we illustrate the parameter region for the multivariate R\'enyi divergence for the particular case of $d=2$. 
The parameter conditions \cref{eq:parameter} are required so that the quantity $D_{\underline{\alpha}}(\vec P_X)$ appropriately behaves as a ``divergence''.
In fact, for any $\underline{\alpha}$ within the allowed parameter region, the quantity $D_{\underline{\alpha}}(\vec P_X)=D_{\underline{\alpha}}(p_X^{(0)},\ldots,p_X^{(d)})$ satisfies the data processing inequality, i.e.,
\begin{equation}\label{eq:DPI}
    D_{\underline{\alpha}}(p_X^{(0)},\ldots,p_X^{(d)})
    \ge D_{\underline{\alpha}}(T(p^{(0)}_X),\ldots,T(p^{(d)}_X))
\end{equation}
holds for any stochastic operator $T$ \cite{farooq2024}.
This can be understood in the sense that operating with $T$ makes the PMFs $p^{(k)}_X$ more similar to each other and the set $P_{\vec{X}}$ of PMFs less informative.
Moreover, it satisfies the positive-definiteness, i.e., $D_{\underline{\alpha}}(\vec P_X)\ge0$ for any $\vec P_X=\big(p^{(0)}_X,\ldots,p^{(d)}_X\big)$ and $D_{\underline{\alpha}}(\vec P_X)=0$ if and only if $p^{(0)}_X=\cdots=p^{(d)}_X$.
In \cref{a:l2-2}, we discuss some monotonicity properties of the divergence $D_{\underline{\alpha}}(\vec P_X)$ with respect to the order parameter $\underline{\alpha}$.
We finally note that the case $d=1$ recovers the (bivariate) R\'enyi divergence of two PMFs $p_X^{(0)}$ and $p_X^{(1)}$ and of order $\alpha \in [0,1) \cup (1,\infty)$ as \cite{renyi, van_Erven_et_al_2014}
\begin{equation}
    \label{eq:Renyi_bi}
    \begin{aligned}
        &D_\alpha(
        p_X^{(0)}
        \|
        p_X^{(1)}
        )
        \\
        &\qquad=
        \frac{1}{\alpha-1}
        \log
        \left[
        \sum_{x\in \mathcal{X}}
        (
        p_X^{(0)}(x)
        )^\alpha 
        (
        p_X^{(1)}(x)
        )^{1-\alpha}
        \right]
    \end{aligned}
\end{equation}
up to the coefficient. 
Furthermore, the case $\alpha=1$ reproduces the Kullback-Leibler divergence (also known as the relative entropy) \cite{KL1951, CT}. 

We now move on to introduce a conditional extension of the multivariate R\'enyi divergence. We assume that, in addition to random variables $X^{(0)},\ldots,X^{(d)}$ with the same alphabet $\mc X$, there is a random variable $G$ on another alphabet $\mc G$ describing side information. 
Taking inspiration from the (bivariate) Bleuler-Lapidoth-Pfister (BLP) conditional R\'enyi divergence \cite{BLP2020}, we introduce a \emph{conditional} multivariate R\'enyi divergence as follows.

\begin{definition}[\emph{Conditional multivariate R\'{e}nyi divergence}]
    \label{def: cond mv Renyi}
    Let $\vec P_{X|G} = \big( p^{(0)}_{X|G}, \ldots, p^{(d)}_{X|G} \big)$ be $(d+1)$ conditional PMFs with respect to a PMF $p_G$ with full support.
    The conditional multivariate R\'{e}nyi divergence of $P_{\vec{X}|G}$ and of orders $(\underline{\alpha},\beta)$, with $\underline{\alpha} = (\alpha_0, \ldots, \alpha_{d}) \in \R^{d+1}$ satisfying $\sum_{k=0}^{d} \alpha_k = 1$ and either (i) or (ii) of \cref{eq:parameter} and $\beta\in \R_{{\scriptscriptstyle \ge0}}$, is denoted as $D_{\underline{\alpha},\beta}(\vec P_{X|G}| p_G)$ and given by
    \begin{equation}
        \label{eq:BLPform}
        \begin{aligned}
            &D_{\underline{\alpha},\beta}(\vec P_{X|G}\big| p_G)\\
            &\coloneqq \!
            \frac{\beta}{\alpha_{\star}\!-\!1}\!\log\!
            \left[
            \sum_{g\in \mathcal{G}}
            p_G(g)\!\!
            \left(
            \sum_{x\in \mathcal{X}}
            \!
            \prod_{k=0}^{d}
            \!\! 
            \big(
            p^{(k)}_{X|G}(x|g)
            \big)^{\!\alpha_k}\!\!
            \right)^{\!\!\!1/\beta}
            \right],
        \end{aligned}
    \end{equation}
    where $\alpha_{\star} \coloneqq \max_{0\leq k\leq d}\alpha_k$ and each $p^{(k)}_{X|G}(\cdot|g)$ is a PMF on $X$ ($\forall g\in\mc G,~ k=0,\ldots,d$). 
\end{definition}

We note that the case $d=1$ with $\beta = \alpha_{\star}$ recovers the (bivariate) conditional R\'enyi divergence of order $\alpha \in (-\infty,0)\cup(0,1)\cup(1,\infty)$ introduced by Bleuler-Lapidoth-Pfister (BLP) \cite{BLP2020} up to the coefficient:
\begin{equation*}
    \label{eq:BLP_bi}
    {\small
        \begin{aligned}
            &
            D^{\rm BLP}_\alpha
            (
            p^{(0)}_{X|G}
            ||
            p^{(1)}_{X|G}
            \big|
            p_G
            )
            \\
            &=
            \frac{\alpha}{\alpha-1}
            \log\!
            \left[
            \sum_{g\in\mc G}
            p_G(g)\!
            \left(
            \sum_{x\in \mc X}
            (p^{(0)}_X\!(x|g))^\alpha
            (p^{(1)}_X\!(x|g))^{1-\alpha}\!
            \right)^{\!\!\!1/\alpha}	
            \right]
            .
        \end{aligned}
    }
\end{equation*}

To develop another information-theoretic understanding of the conditional multivariate R\'{e}nyi divergence, we now address data processing inequalities that the quantity satisfies.
To our knowledge, there is not yet a commonly accepted definition for data processing for conditional PMFs, so establishing general data processing inequalities for the conditional divergence is problematic. 
However, in \cite{BLP2020} some steps towards this were taken in the bivariate case, and let us generalise them to our multivariate case \cref{eq:BLPform}.
The first inequality is about postprocessing the PMFs $(p^{(0)}_{X|G}, \ldots, p^{(d)}_{X|G})$.
\begin{prop}[\emph{Data processing inequality with respect to the main system}]
    \label{l:l1}
    Consider conditional PMFs $\vec P_{X}=\big( p^{(0)}_{X|G}, \ldots, p^{(d)}_{X|G} \big)$ and the conditional multivariate R\'{e}nyi divergence
    $D_{\underline{\alpha},\beta}(P_{\vec{X}|G}\big|p_G)$ in \cref{eq:BLPform}. 
    Let $T_{Y|XG}$ be a stochastic operator taking each conditional PMF $p_{X|G}^{(k)}$ on $\mc X\times \mc G$ into a conditional PMF $q_{Y|G}^{(k)}:=T_{Y|XG}(p_{X|G}^{(k)})$ on $\mc Y\times \mc G$ $(k=0,\ldots,d)$ by
    \begin{equation*}\label{}
        q^{(k)}_{Y|G}(y|g)=\sum_{x\in \mc X}t_{Y|XG}(y|x,g)p^{(k)}_{X|G}(x|g),
    \end{equation*}
    where $t_{Y|XG}(\cdot|x,g)$ is the conditional PMF on $\mc Y$ defined by $T_{Y|XG}$ ($\forall x\in \mc X, g\in\mc G$).
    Then
    \begin{equation*}
        \begin{aligned}
            D_{\underline{\alpha},\beta}
            \big(
            p^{(0)}_{X|G}, \ldots, p^{(d)}_{X|G}
            \big|
            p_G
            \big)
            &\ge
            D_{\underline{\alpha},\beta}
            \big(
            q^{(0)}_{Y|G}, \ldots, q^{(d)}_{Y|G}
            \big|
            p_G
            \big)
        \end{aligned}
    \end{equation*}
    holds.
\end{prop}
The proof follows immediately from the data processing inequality for the unconditional $D_{\underline{\alpha}}$ in \cref{eq:DPI} and the form of \cref{eq:BLPform} for the multivariate conditional divergence.
We now consider postprocessing of $p_G$. 
\begin{prop}[\emph{Data processing inequality with respect to the conditioning system}]\label{l:l2}
    Consider conditional PMFs $\vec P_{X|G}=\big( p^{(0)}_{X|G}, \ldots, p^{(d)}_{X|G} \big)$ and assume $p_{X|G}^{(k)} = r_X^{(k)}$ ($k =1,\ldots,d$) to be independent of $G$.
    Through a stochastic operator $T_{H|G}$ and its pseudo-inverse $T^\dagger_{G|H}$, we define a PMF $q_H^{}$ on $\mc H$ and a conditional PMF $q_{X|H}^{(0)}$ on $\mc X\times \mc H$ respectively by
    \begin{align*}\label{}
        &q_{H}(h)=\sum_{g\in \mc G}t_{H|G}(h|g)p_{G}(g),\\
        &
        \begin{aligned}
            q^{(0)}_{X|H}(x|h)&=\sum_{g\in \mc G}t^\dagger_{G|H}(g|h)p^{(0)}_{X|G}(x|g)\\
            &=\sum_{g\in \mc G}\frac{p_G(g)}{q_H(h)}t_{H|G}(h|g)p^{(0)}_{X|G}(x|g),
        \end{aligned}
    \end{align*}
    where $t_{H|G}$ and $t^\dagger_{H|G}$ are the conditional PMFs associated respectively with $T_{H|G}$ and $T^\dagger_{G|H}$.
    Then, with $\alpha_0=\max_k\alpha_k$ and $\beta=\alpha_0$, the multivariate conditional R\'enyi divergence $D_{\underline{\alpha},\alpha_0}(\vec P_{X|G}\big|p_G)$ satisfies
    \begin{equation}
        \label{eq:BLPineq}	
        \begin{aligned}
            &D_{\underline{\alpha}, \alpha_0}
            \big(
            p^{(0)}_{X|G}, 
            r_X^{(1)}, 
            \ldots, 
            r_X^{(d)}
            \big|
            p_G
            \big)
            \\ 
            &\qquad\qquad \geq 
            D_{\underline{\alpha}, \alpha_0}
            \big(
            q^{(0)}_{X|H}, 
            r_X^{(1)}, 
            \ldots, 
            r_X^{(d)}
            \big|
            q_H
            \big)
            .
        \end{aligned}
    \end{equation}
\end{prop}
The proof of this statement given in \cref{a:l2}. 
Suppose above that the stochastic operator $T_{H|G}$ does not depend on $g\in\mc G$, i.e., $t_{H|G}(h|g)=q_H(h)$ ($\forall g\in \mc G, h\in\mc H$) holds.
It implies $q^{(0)}_{X|H}=p_X^{(0)}$ and thus an important corollary follows.
\begin{corollary} [\emph{Data processing inequality}]
    \label{c:c1}
    \begin{equation*}
        \label{eq:BLPineq2}
        \begin{aligned}	
            D_{\underline{\alpha},\alpha_0}
            \big(
            p^{(0)}_{X|G},r_X^{(1)}, \ldots,r_X^{(d)}\big|p^{}_G
            \big)
            \ge
            D_{\underline{\alpha}}
            \big(
            p^{(0)}_{X},r_X^{(1)}, \ldots, r_X^{(d)}
            \big)
            .
        \end{aligned}
    \end{equation*}
\end{corollary}
This last data processing inequality will play an important role in the subsequent sections, as it explicitly relates the conditional multivariate R\'enyi divergence to its unconditional counterpart. 
This will prove useful when providing an operational interpretation of multivariate R\'enyi divergences in terms of betting games with multiple lotteries. 
In order to do so, let us now address the theory of expected utility from the economic sciences.

\section{Expected utility theory}
\label{s:s3}
	
\begin{figure}[h!]
    \centering
    \includegraphics[scale=1.57]{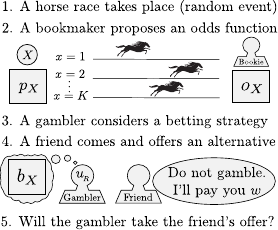}
    \caption{
        Illustration of a single-lottery betting game. A random event (horse race) is described by a random variable $X$ and distributed according to a probability mass function $p_X$. A bookmaker proposes an odds function $o_X$, and a gambler considers placing a betting strategy $b_X$. The gambler is assumed to have a risk-averse character represented by a utility function $u_{R}$, with $R$ the risk-averse parameter. Before the gambler fully committing to playing the game, a friend proposes to pay the gambler a fixed amount, say $w$, to persuade him to walk away from the betting game. Will the gambler take the friend's offer over gambling on the lottery? The solution to this decision problem is given by the isoelastic certainty equivalent $w^{\rm ICE}_{R} \equiv w^{\rm ICE}(p_{X}, o_X, b_{X}, u_{R})$ of the betting game as it represents the economic-theoretic value that the gambler assigns to the lottery in question. Explicitly, the gambler will choose to gamble whenever $w^{\rm ICE}_{R} > w$, and will take the friend's offer otherwise. 
    }
    \label{fig:fig2}
\end{figure}

We begin by describing the operational task of horse betting games (or simply betting games) \cite{CT, LN_Moser, BLP2020, kelly}. 
Consider three types of agents: a referee, a gambler, and a bookmaker (see \cref{fig:fig2}). 
The referee is in charge of a probabilistic event governed by a PMF $p_X$ on an outcome set $\mc X$. 
We interpret the event as a horse race; the value $p_X(x)$ exhibits the probability for each horse $x\in \mc X$ to win the race. 
The bookmaker is in charge of proposing a non-negative function $o_X\colon \mc X\to\R_{{\scriptscriptstyle \ge0}}$, the ``odds" (or ``payoff'') function, on $\mc X$ and announces it to the gambler. 
Under this setting, the gambler is going  to ``bet" on every possible horse $x\in \mc X$. 
In other words, the gambler can say, for instance: ``I believe horse 1 will win with probability $20\%$" and ``horse 2 with probability $30\%$", and similarly with the rest of $\mc X$. 
This can be modelled by asking the gambler to provide a \emph{betting strategy} in the form of a PMF $b_X$ on $\mc X$ ($b_X(1)=0.2, b_X(2)=0.3,...$ in the previous example).
In a standard scenario of economic theory, the amount of \textit{wealth} (money, goods, etc.) is represented by positive real numbers. 
Naturally, we interpret the amount $w_1 \in \R_{{\scriptscriptstyle \ge0}}$ as ``more valuable'' than $w_2\in\R_{{\scriptscriptstyle \ge0}}$ if $w_1>w_2$. 
Now let the gambler start with an initial amount of wealth $w^i > 0$.
The gambler is assumed to distribute all the initial wealth as bets according to $b_X$, i.e., the gambler bets $b_X(x)w^i$ on each horse $x$.
After the gambler defines the betting strategy $b_X$, the referee holds the race and reveals the winning horse, say $x'$.
The bookmaker then rewards the gambler with an amount of wealth (final wealth) proportional to the betting done on the correct index given as $w^f =  b_X(x')o_X(x') w^i$ (and nothing otherwise), with $o_X$ the pre-established odds function. 
From this point forward, and for simplicity, we assume that the gambler starts with initial wealth given by a unit of wealth $w^i=1$, and so final wealth is simply $w^f=w_X(x') = b_X(x')o_X(x')$. 
One important property for later discussion is the \textit{fairness} of the odds, which is given by $F \coloneqq \sum_x 1/o_X(x)$. The odds are considered to be fair when $F=1$, sub-fair when $F>1$, and super-fair when $F<1$ \cite{CT, LN_Moser, BLP2020, kelly}.

Let us take these elements a step further from the perspective of \textit{expected utility theory} (EUT), arguably the main theoretical framework used in economics and finance since the 1940s \cite{risk_vNM}. 
A key ingredient of EUT is the notion of \textit{utility functions}. 
A utility function $u\colon\R\to\R$ expresses the level of ``satisfaction'' of a rational agent when acquiring an amount of wealth; the value $u(w)$ quantifies how satisfied the agent is when receiving the amount of wealth $w$. 
For simplicity, in this paper we mainly study the case when wealth $w$ and utility $u$ are non-negative, while the converse situations express ``loss'' and ``dissatisfaction'' respectively. 
A natural requirement is that the utility function $u$ is monotonically increasing with wealth, i.e., $u(w_1)>u(w_2)$ if $w_1>w_2$, simply meaning that a larger amount of wealth leads to a larger satisfaction for the agent.
The principle of EUT is that under uncertain situations, rational agents make decisions based not on expected wealth but on expected utility.
A specific context in which this principle plays a crucial role is the decision problem  involving uncertain gains such as \emph{lotteries}.

The main subject of this paper is the notion of lotteries in EUT. 
A lottery is a pair $(p_X,w_X)$ of a PMF $p_X$ on an outcome set $\mc X$ and a wealth (or reward) function $w_X$, describing the situation where wealth $\{w_X(x)\}_{x \in \mathcal{X}}$ is randomly distributed according to $\{p_X(x)\}_{x \in \mathcal{X}}$. 
It is clear that the setup of a betting game described before corresponds to a lottery with a wealth function given by $w_X = b_X o_X$ and the PMF $p_{X}$. 
Important to our discussion later on, here we note that this setup involves a \emph{single}-commodity lottery. 
Economists have derived ways of quantifying how ``attractive" lotteries are to rational agents. 
In short, the economic-theoretic ``value" of a lottery $(p_X,w_X)$ for an agent (a gambler) modelled by a utility function $u$ is given by the \textit{certainty equivalent} (CE) $w^{\rm CE} \equiv w^{\rm CE}(p_{X},w_X,u)$, which is defined as the amount of wealth that generates the same level of satisfaction as the expected utility of the lottery, and so is given by
\begin{align}
    u(
    w^{\rm CE}
    )
    \coloneqq
    \mathbb{E}_{p_{X}}
    \left[
    u(w_X)
    \right]
    =
    \sum_{x\in \mc X}
    p_X(x)
    u(w_X(x))
    .
\end{align}
This definition indicates that the expected (uncertain) profit of the lottery is ``equivalent'' to the certain gain of $w^{\rm CE}$ for the agent. 
That is, according to the principle of EUT, the agent chooses the lottery instead of a fixed amount of wealth, say $w$, whenever $w^{\rm CE} > w$. In this sense, the quantity $w^{\rm CE}$ can be considered as a reasonable measure of the economic-theoretic value that the agent assigns to the lottery. From the perspective of economics, the certainty equivalent also expresses the agent's attitude towards \textit{risk}. To see this, suppose that $w^{\rm CE}<\mathbb{E}_{ p_{X}}[w_X]$. In this case, although the agent is expected to receive a larger amount of wealth through the gamble, the agent is satisfied enough by certainly gaining $w^{\rm CE}$, and so this describes that the agent is \textit{risk averse}. With similar arguments, when $w^{\rm CE}>\mathbb{E}_{ p_{X}}[w_X]$, the agent is seen as \textit{risk seeking}, and when $w^{\rm CE}=\mathbb{E}_{p_{X}}[w_X]$, the agent is considered \textit{risk neutral} \cite{risk_bernoulli, risk_arrow, risk_pratt}.


The certainty equivalent $w^{\mathrm{CE}}$ depends on the lottery in question as well as the utility function $u$. 
There are several utility functions of interest to economists, one of which is the \emph{isoelastic utility function} 
\begin{equation}
    \label{eq:IEU}
    u_R(w)
    \coloneqq
    \frac{w^{1-R}-1}{1-R}
    \quad
    (R\in \R).
\end{equation}
The parameter $R$ describes the agent's attitude of risk aversion through the degree of convexity (or concavity) of the function \cite{QBG_2022};
$R>0$ describes a risk-averse attitude, $R<0$ risk-seeking, and $R=0$ risk neutral. 
Below we mainly consider risk-averse agents and use a simpler form \cite{Recursive, Meyer2014}
\begin{equation}
    \label{eq:IEU2}
    u_R(w)
    \coloneqq
    \frac{w^{1-R}}{1-R}
    \quad
    (R\in \R_{{\scriptscriptstyle \ge0}}).
\end{equation}
Later we will see that the assumption of risk aversion is related to the data processing inequalities for multivariate R\'enyi divergences.
Note that the function $u_R$ is set as $u_R(w)=\log w$ for $R=1$ by taking the limit in \cref{eq:IEU}.
In the following, the diverging factor $(1-R)^{-1}$ is usually cancelled, and thus the case $R=1$ is treated by setting $R\to 1$ at the final stage of the calculation.
The certainty equivalent of the isoelastic function, or simply the \textit{isoelastic certainty equivalent} (ICE), for a betting game is then given by
\begin{align*}
    w^{\rm ICE}_R
    =
    \left(
    \sum_{x\in\mc X}
    p_X(x)
    \big(
    b_X(x)
    o_X(x)
    \big)^{1-R}
    \right)^\frac{1}{1-R}.
\end{align*}
The notion of the ICE will play a crucial role in the following analysis. This finishes the description of betting games and EUT with \textit{single}-commodity lotteries. Let us move on to the new scenario of lotteries with \textit{multiple} commodities that we introduce in this work.

	
\section{Betting games with multiple lotteries}
\label{s:s4a}
\begin{figure}
    \includegraphics[scale=1.57]{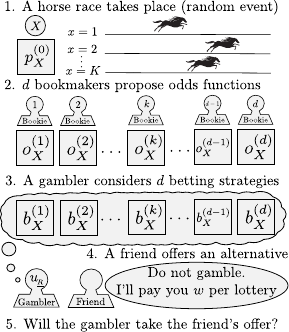}
		\caption{
			Schematic representation of a betting game extended to a multi-lottery framework. 
			A random event (horse race) is described by a random variable $X$ distributed according to a probability mass function $p_X^{(0)}$. In this setting, a group of $d$ bookmakers propose $d$ odds functions $\vec{O}_X=(o_X^{(1)},\ldots,o_X^{(d)})$, and a gambler places $d$ betting strategies $\vec{B}_X=(b_X^{(1)},\ldots,b_X^{(d)})$. Accordingly, the gambler's risk profile is characterised by a risk-aversion vector $\underline{R} = (R_1,\ldots,R_d)\in\mathbb{R}_{\ge 0}^d$ within the utility function $u_{\underline{R}}$, where each $R_k$ is the risk-aversion parameter associated with the $k$th lottery. 
			To persuade the gambler to withdraw from the betting, a fixed offer of $w$ per lottery is presented. The decision to accept this offer or play depends on the isoelastic certainty equivalent $w^{\mathrm{ICE}}_{\underline{R}} \equiv w^{\mathrm{ICE}}(p_{X}^{(0)},\vec{O}_X,\vec{B}_{X},u_{\underline{R}})$, which quantifies the gambler's valuation of the lotteries. Explicitly, the gambler chooses to play (rejecting the persuasion) whenever $w^{\mathrm{ICE}}_{\underline{R}} > w$.
			}
    \label{fig:uncond}
\end{figure}
Expected utility theory can accommodate the analysis of multi-commodity wealth $\vec w = (w^{(1)}, \ldots, w^{(d)})\in\R^{d}$ by considering utility functions $u:\R^{d} \to \R$ \cite{MC1, MC2, MC3, MC4}. Similarly to the single-commodity case, here $u(\vec w_{})$ represents the level of satisfaction that an agent experiences when acquiring the vector of commodities $\vec w_{}\in\R^{d}$. 

This situation can be combined with betting games to be rephrased as \textit{betting games with multiple lotteries} or simply \textit{multi-lottery betting games}. 
Specifically, in a $d$-lottery betting game, played by a referee, a gambler, and $d$ bookmakers, the referee is in charge of a random event (a horse race) governed by a PMF $p_X^{(0)}$ on $\mc X$. 
Unlike the single-commodity case, here the gambler is rewarded with a set of \emph{multiple} commodities described by wealth vectors $\vec w =(w^{(1)}, \ldots, w^{(d)})$ from \emph{multiple} bookmakers. 
That is, there are $d$ bookmakers offering $d$ odds functions $\vec O_X = (o_X^{(1)}, \ldots, o_X^{(d)})$ in advance and the gambler placing $d$ bets $\vec B_{X} = (b^{(1)}_X, \ldots, b^{(d)}_X)$ of PMFs, where $o_X^{(k)}\colon \mc X\to\R_{{\scriptscriptstyle \ge 0}}$ and $b_X^{(k)}\colon \mc X\to[0,1]$ are respectively the odds and betting strategy for the $k$th component $(k=1,\ldots,d)$.
The referee then reveals the winning label (horse), say $x'$, and asks the bookmakers to reward the gambler with a \emph{multi-commodity wealth} vector $\vec W(x') = (w_X^{(1)}(x'), \ldots,w_X^{(d)}(x'))$ with each $w_X^{(k)}(x') = b^{(k)}_{X}(x') o_X^{(k)}(x')$ $(k =1,...,d)$ a wealth function on $\mc X$.
This scenario can be effectively understood as the gambler playing \emph{$d$ lotteries} $\{(p_X^{(0)},w_X^{(k)})\}_{k=1}^d\equiv(p_X^{(0)},\vec{W})\equiv(p_X^{(0)},\vec O_X,\vec{B}_X)$.
As in the previous construction, we are interested in the economic-theoretic ``value" that a rational agent with a utility function $u:\R^{d} \to \R$ assigns to the $d$ lotteries $(p_X^{(0)},\vec{W})$. 
One possible approach is to consider the certainty equivalent (CE) vector $\vec w^{\rm CE}\coloneqq (w^{\,\rm CE},w^{\,\rm CE},...,w^{\,\rm CE})\in\R^{d}$ defined via
\begin{align}
    u
    (
    \vec w^{\,\rm CE}
    )
    \coloneqq
    \mathbb{E}_{p_{X}}
    \big[
    u(
    \vec W
    )
    \big]
    =
    \sum_{x\in \mathcal{X}}
    p_X(x)
    u(
    \vec W(x)
    )
    .\label{eq:multiCE}
\end{align}
Similarly to the single-commodity scenario, here the CE $w^{\,\rm CE}(p_X^{(0)},\vec W,u)\equiv w^{\rm CE}$ represents the amount of \emph{certain} per-commodity wealth that generates the same level of satisfaction for the gambler as playing the $d$ lotteries $(p_X^{(0)},\vec{W})$. 
In other words, the gambler is willing to accept a \emph{certain} amount of wealth $w^{\,\rm CE}$ per lottery (so $dw^{\,\rm CE}$ in total), instead of playing the $d$ lotteries $(p_X^{(0)},\vec{W})=\{(p_X^{(0)},w_X^{(k)})\}_k$ to obtain \emph{uncertain} rewards $\vec W(x')=(w_X^{(1)}(x'),\cdots, w_X^{(d)}(x'))$.
There are several potential utility functions of interest, one of which is the \emph{multi-commodity isoelastic utility function} $u_{\underline{R}}\colon\R^d\to\R$ generalising \cref{eq:IEU2} as \cite{Escobar-Anel_2022, EscJiao2024}
\begin{align}
    u_{\underline{R}}
    (\vec w)
    \coloneqq
    \prod_{k=1}^{d}
    \frac{(w^{(k)})^{1-R_k}}{1-R_k}
    ,
    \label{eq:multi-IEU}
\end{align}
where $\underline{R} =(R_1, \ldots, R_d)\in\R_{{\scriptscriptstyle \ge0}}^{d}$ is called the \emph{risk-aversion vector} implying that the gambler behaves in a risk-averse manner for every lottery $k=1,\ldots,d$. 
This multi-commodity utility function satisfies various desirable properties expected in the multi-commodity setting \cite{EA_2023}.
When applied to our betting game, it particularly gives the isoelastic certainty equivalent (ICE) $w^{\mathrm{ICE}}_{\underline{R}}(p_{X}^{(0)},\vec O_X,\vec{B}_{X},u_{\underline{R}})$ by
\begin{equation}\label{eq:multiICE}
    \begin{aligned}
        &w^{\mathrm{ICE}}_{\underline{R}}(p_{X}^{(0)},\vec O_X,\vec{B}_{X},u_{\underline{R}})\\
        &=\left(
        \sum_{x\in\mc X}
        p_X^{(0)}\!(x)
        \!\prod_{k=1}^{d}\!
        \big(
        b_X^{(k)}(x)
        o_X^{(k)}(x)
        \big)^{\!1-R_k}\!
        \right)^{\! \frac{1}{\sum\limits_{k=1}^d\!(1-R_k)}}
    \end{aligned}
\end{equation}
through the formula \cref{eq:multiCE}.
In \cref{fig:uncond}, we further illustrate the task of multi-lottery betting games explained so far.

	
\section{Unconditional multivariate R\'enyi divergence and multi-lottery betting games}
\label{s:s4r1}
We are now in position to derive a connection between the certainty equivalent of multi-lottery betting games and the multivariate R\'{e}nyi divergence.
\begin{result}[\emph{Economic-theoretic interpretation of the unconditional multivariate R\'enyi divergence}]
    \label{r:r1}
    Consider a $d$-lottery betting game involving a random event (horse race), $d$ bookmakers ($d \geq 1$), and a gambler. 
    The random event is described by a PMF $p_X^{(0)}$, and the $d$ bookmakers offer $d$ odds $\vec O_X = (o_X^{(1)}, \ldots, o_X^{(d)})$. 
    The gambler then proceeds to place $d$ betting strategies described by PMFs $\vec B_{X} = (b^{(1)}_X, \ldots, b^{(d)}_X)$, thus effectively establishing $d$ lotteries as $\{(p_X^{(0)},w_X^{(k)})\}_{k=1}^d\equiv (p_X^{(0)},\vec O_X,\vec{B}_X)$ with $w_X^{(k)} = b_X^{(k)} o_X^{(k)}$ $(k=1,\ldots,d)$.
    The gambler is also assumed to behave according to a multi-commodity isoelastic utility function $u_{\underline{R}}$, where $\underline{R} = (R_1,\ldots,R_d)$ is a risk-aversion vector with each risk-aversion value $R_k \in \R_{{\scriptscriptstyle \ge 0}}$ specific to lottery $k$ and satisfying either of the following conditions:
    \begin{equation}
        \label{eq:ra_condition}
            \begin{aligned}
                \mathrm{(i)} &\ \mbox{$R_k\geq1$ 
                    $(k=1, \ldots, d)$, }\\
                \mathrm{(ii)} &\ \mbox{$0< R_k< 1$ $(k=1, \ldots, d)$ and 
                    $\sum\limits_{k=1}^{d}\!R_k>d-1$}.
            \end{aligned}
    \end{equation}
    With these elements in place, the economic-theoretic value that the gambler assigns to the $d$ lotteries in question is given, according to expected utility theory, by the isoelastic certainty equivalent (ICE) $w^{\rm ICE}_{\underline{R}} \equiv w^{\rm ICE}(p_X^{(0)}, \vec O_X, \vec{B}_X, u_{\underline{R}})$ via \cref{eq:multiICE}. 
    This ICE can be decomposed as
    \begin{align}
        \label{eq:wCEform}
        &\log
        w^{\rm ICE}(p_X^{(0)}, \vec O_X, \vec{B}_X, u_{\underline{R}})
        =
        D_{\underline{\alpha}}(p_X^{(0)}, p_X^{(1)}, \ldots, p_X^{(d)})
        \nonumber
        \\
        &+
        \sum_{k=1}^{d}
        \frac{\alpha_k}{\alpha_0-1}
        \left(
        D_{S_{k}}
        \big(
        q_X^{(k)}
        \big\|
        b^{(k)}_X
        \big)
        +
        \log
        F^{(k)}
        \right)
        ,
    \end{align}
    where $D_{\underline{\alpha}}(p_X^{(0)}, p_X^{(1)}, \ldots, p_X^{(d)})$ is the multivariate R\'enyi divergence of PMFs $p_X^{(0)}$ and
    \begin{equation}
        \label{eq:E_ge1}
        p_X^{(k)}
        \coloneqq
        \frac{1}{
            F^{(k)}o_X^{(k)}
        },
        \ 
        F^{(k)}
        \coloneqq
        \sum_{x\in \mc X}
        \frac{1}{o_X^{(k)}(x)}
        ,
    \end{equation}
    and $D_{S_k}(q_X^{(k)}\|b_X^{(k)})$ is the bivariate R\'enyi divergence \cref{eq:Renyi_bi} of $b_X^{(k)}$ and a PMF $q_X^{(k)}$ independent of $b_X^{(k)}$ $(k=1, \ldots, d)$.
    The explicit expression of $q_X^{(k)}$ is detailed in \cref{a:r1}.
    The orders $\underline{\alpha} = (\alpha_0, \ldots, \alpha_{d})$ and $S_k$ for these divergences are determined by the risk-aversion vector \cref{eq:ra_condition} and given respectively as
    \begin{equation}
        \label{eq:ra_alpha}
        \alpha_0
        \coloneqq\left(1+\sum_{k=1}^{d}(R_k-1)\right)^{-1}\!\!\!, \ \  \alpha_k\coloneqq(R_k-1) \alpha_0,
    \end{equation}
    and
    \begin{align}
        S_k
        \coloneqq 
        \frac{
            \sum_{l=0}^k \alpha_{l}
        }{
            \sum_{l=0}^{k-1} \alpha_{l}
        }\ \ (\ge0).
    \end{align}
\end{result}

The full proof of this statement is in \cref{a:r1}, where we also developed a continuous extension of the result to general outcome sets.
There are several aspects of this result worth analysing. First, this result is effectively a mathematical identity and thus holds very generally for all PMFs $p_X^{(0)}$, all types of odds $\vec O_X$ proposed by the bookmakers, all types of betting strategies $\vec{B}_X$ placed by the gambler, as well as for arbitrary agents with general attitudes towards risk described by the multi-commodity utility function $u_{\underline{R}}$ as in \cref{eq:multi-IEU}. 
Second, it is worth highlighting that whilst the left-hand side in \cref{eq:wCEform} corresponds to a purely economic-theoretically operational task, specifically the certainty equivalent of $d$ lotteries, the right-hand side contains information-theoretic quantities in the form of a multivariate R\'enyi divergence as well as $d$ bivariate R\'enyi divergences.
This result thus establishes a direct link between notions in the realm of the theory of games and economic behaviour on the one hand, and information-theoretic quantities on the other.
It is also of particular interest that the quantitative characterisation of the gambler's risk-averse attitude for each lottery completely reproduces the order restrictions for the multivariate R\'enyi divergence.
To see this, let us remark that the conditions \cref{eq:ra_condition} and \cref{eq:ra_alpha} for the risk-aversion vector are equivalent to \cref{eq:parameter} and $R_k=1+\alpha_k/\alpha_0$.
The former constraints display an information-theoretically clear interpretation; the gambler treats every lottery $k$ with a similar degree of risk aversion: $R_k\ge1$ (sensitive to risk) or $0<R_k<1$ (insensitive to risk) for every $k$.
In other words, the seemingly peculiar conditions for $\underline{\alpha}$ can now be understood as, not just mathematical requirements that make $D_{\underline{\alpha}}$ information-theoretically consistent, but also expressing the gambler's economic-theoretic consistency towards risk in a multi-lottery betting game, where the gambler never prioritises a specific lottery and acts analogously for all lotteries.
Finally, the decomposition in the right-hand side of \cref{eq:wCEform} can be divided into three terms: a first term concerning the multivariate R\'enyi divergence of PMFs $p_X^{(k)}$, a second term concerning the bivariate R\'enyi divergences, and a third term containing the coefficients $\{F^{(k)}\}_k$.
These terms are rich in explanatory power, which we exploit next.

Let us start by analysing the term in \cref{eq:wCEform} containing the $d$ bivariate R\'enyi divergences. 
The coefficient $\alpha_k/(\alpha_0-1)$ is non-positive, and taking into account that the bivariate R\'enyi divergence $D_{S_{k}}(q_X^{(k)}\|b^{(k)}_X)$ is non-negative, this makes these terms non-positive. 
We can use this decomposition to extract the optimal betting strategies for the maximisation of the certainty equivalent and, consequently, the term can be interpreted as the penalisation to the gambler's satisfaction for not implementing the optimal bets. Explicitly, the term $D_{S_{k}} \big( q_X^{(k)} \big \| b^{(k)}_X)$ is the gap between the gambler's bet $b_X^{(k)}$ and the optimal bet $q_X^{(k)}$ for lottery $k$. We summarise this in the following corollary.
\begin{corollary}[\emph{Optimal betting strategies}] \label{c:c2}
    The log isoelastic certainty equivalent of a $d$-lottery betting game with a gambler implementing optimal betting strategies reads
    \begin{align*}
        &
        \max_{\vec{B}_X}~
        \log
        w^{\rm ICE}(p_X^{(0)}, \vec O_X, \vec{B}_X, u_{\underline{R}})
        \\
        &\qquad\qquad=
        D_{\underline{\alpha}}
        (\vec{P}_X)
        +
        \sum_{k=1}^{d}
        \frac{\alpha_k}{\alpha_0-1}
        \log
        F^{(k)}
        ,
        \nonumber
    \end{align*}
    where the optimisation is taken over all betting strategies represented by PMFs $\vec B_X$. 
\end{corollary}

We continue the analysis of the decomposition in \cref{eq:wCEform} and consider now the term containing the constants $F^{(k)} = \sum_x 1/o_X^{(k)}(x)$. This quantity depends exclusively on the odds proposed by the bookmakers and, more specifically, describes the fairness of the odds. 
When all the odds are sub-fair ($F^{(k)} > 1$, $\forall k$), the value of the lottery decreases (taking into account the coefficient $\alpha_k/(\alpha_0-1)$ is non-positive), as one can intuitively expect. Similarly, when all the odds are super-fair ($F^{(k)} < 1$, $\forall k$), the value of the lotteries on the other hand increases, which rarely happens as bookmakers usually offer odds that are sub-fair.
Finally, when all the odds are fair ($F^{(k)} = 1$, $\forall k$), these terms disappear.
Additionally, with this restriction, we obtain $p^{(k)}_X = (o_X^{(k)})^{-1}$ ($k=1, \ldots, d$) as per \cref{eq:E_ge1}. 
Writing $d$ fair odds functions as $\vec O_X^{\mathrm{Fair}}$, we can summarise the latter case in the following corollary.
\begin{corollary}[\emph{The unconditional multivariate R\'enyi divergence characterises multi-lottery betting, optimal betting strategies, and fair odds}] \label{c:c3}
    The log isoelastic certainty equivalent of a $d$-lottery betting game with fair odds 
    and a gambler implementing optimal betting strategies
    reads
    \begin{align*}
        \underset{\vec{B}_X}{\max}
        ~
        \log
        w^{\rm ICE}(p_X^{(0)}, \vec O_X^{\mathrm{Fair}}, \vec{B}_X, u_{\underline{R}})
        =
        D_{\underline{\alpha}}
        (\vec{P}_X)
        ,
    \end{align*}
    where the optimisation is taken over all betting strategies.
\end{corollary}

This latter statement provides a conceptually clean interpretation of the multivariate R\'{e}nyi divergence as the economic-theoretic value that rational agents assign to multi-lottery betting games, when allowed to maximise over betting strategies, and when all the odds are fair.  
In particular, one appealing feature of this statement is that the R\'enyi orders $\alpha_k$ can now also be interpreted, as they explicitly represent the degree of risk aversion manifested by the agent towards each lottery in question as $R_k = 1+\alpha_{k}/\alpha_0$ ($k=1,\dots,d$).
In addition to being of purely mathematical interest, the multivariate R\'enyi divergence is now endowed with an economic-theoretic interpretation. 
In order to reinforce these conceptual aspects, we will further elaborate on this idea in subsequent sections by introducing tasks in the framework of resource theories within general probabilistic theories, where these results will find further applicability. We now move on to consider a generalised scenario where the rational agent has access to side information. 
	
	\section{Conditional multivariate R\'enyi divergence and multi-lottery betting games}
	\label{s:s4r2}
\begin{figure}
		\includegraphics[scale=1.57]{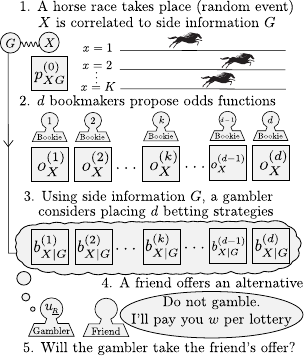}
		\caption{
			Multi-lottery betting games with side information. Unlike standard multi-lottery betting games described in \cref{fig:fig2}, the random variable $X$ is now correlated to another random variable $G$ and jointly distributed according to a joint probability mass function $p_{XG}^{(0)}$. The gambler places $d$ betting strategies  $\vec{B}_{X|G}=(b_{X|G}^{(1)}, \ldots, b_{X|G}^{(d)})$ having access to the side information $G$. The economic-theoretic value of the lotteries is given by $w^{\rm ICE}_{\underline{R}} \equiv w^{\rm ICE}(p_{XG}^{(0)},\vec O_X,\vec{B}_{X|G},u_{\underline{R}})$.
		}
		\label{fig:cond}
	\end{figure}

	We now consider multi-lottery betting games where the gambler can place betting strategies after having access to side information. 
	In \cref{fig:cond}, we illustrate this scenario and fully describe it as follows.

	We consider a $d$-lottery game constituted almost in the same way as before; a random event (a horse race) takes place according to a PMF $p_X^{(0)}$ on an outcome set $\mc X$, $d$ bookmakers offer $d$ odds functions $\vec O_X=(o_X^{(1)}, \ldots, o_X^{(d)})$, and a gambler makes $d$ bets.
	However, this time we assume that the random variable $X$ is correlated to another random variable $G$ whose outcome set is $\mc G$ and thus there is a joint PMF $p_{XG}^{(0)}$ on $\mc X\times\mc G$.
	Also, we allow the gambler to access $G$ and propose $d$ betting strategies according to its outcome.
	The gambler's bets are now $d$ conditional PMFs $\vec{B}_{X|G}=(b_{X|G}^{(1)},\ldots,b_{X|G}^{(d)})$ on $\mc X\times\mc G$ and yield multi-commodity wealth $(b_{X|G}^{(1)}(x|g)o_X^{(1)}(x),\ldots,b_{X|G}^{(d)}(x|g)o_X^{(d)}(x))$ when the winning label is $X=x$ and the gambler receives $G=g$.
	The scenario is effectively equivalent to the gambler playing $d$ lotteries $\{p_{XG}^{(0)}, b_{X|G}^{(k)}o_X^{(k)}\}_{k=1}^d\equiv (p_{XG}^{(0)},\vec O_X,\vec{B}_{X|G})$, where $b_{X|G}^{(k)}o_X^{(k)}\colon (x,g)\mapsto b_{X|G}^{(k)}(x|g)o_X^{(k)}(x)$ is the $k$th wealth function.
	Then, by means of the formula \cref{eq:multiICE}, the value $w^{\mathrm{ICE}}_{\underline{R}}(p_{XG}^{(0)},\vec O_X,\vec{B}_{X|G},u_{\underline{R}})$ of this $d$-lottery betting game for the gambler modelled by the multi-commodity isoelastic utility function $u_{\underline{R}}$ in \cref{eq:multi-IEU} with a risk-aversion vector $\underline{R}$ is
	\begin{equation}\label{eq:multiICE2}
		{\small
			\begin{aligned}
				&w^{\mathrm{ICE}}_{\underline{R}}(p_{XG}^{(0)},\vec O_X,\vec{B}_{X|G},u_{\underline{R}})\\
				&\!=\!
				\left(
				\sum_{x\in\mc X}\!
				p_{XG}^{(0)}(x,\!g)\!\prod_{k=1}^{d}\!
				\big(
				b_{X|G}^{(k)}(x|g)
				o_X^{(k)}(x)
				\big)^{\!1-R_k}\!\!
				\right)^{\! \frac{1}{\!\sum\limits_{k=1}^d\!(1-R_k)}}\!.
			\end{aligned}
		}
	\end{equation}
	We can now present the conditional counterpart of \cref{r:r1}. 
	\begin{result}[\emph{Economic-theoretic interpretation of the conditional multivariate R\'enyi divergence}]
		\label{r:r2}
		Consider a $d$-lottery betting game involving a random event, $d$ bookmakers ($d \geq 1$), and a gambler. 
		The random event is expressed by a random variable $X$ and is correlated to side information $G$ to be jointly described by the PMF $p^{(0)}_{XG}$. 
		The $d$ bookmakers offer $d$ odds $\vec O_X = (o_X^{(1)}, \ldots, o_X^{(d)})$. 
		The gambler has access to the side information $G$ and uses this to place $d$ betting strategies by conditional PMFs $\vec B_{X|G} = (b^{(1)}_{X|G}, \ldots, b^{(d)}_{X|G})$, thus effectively establishing $d$ lotteries $\{p^{(0)}_{XG},w_X^{(k)}\}_{k=1}^d\equiv (p_{XG}^{(0)},\vec O_X,\vec{B}_{X|G})$ with $w_X^{(k)}=b_{X|G}^{(k)}o_X^{(k)}$ ($k=1,\ldots,d$).
		The gambler is also assumed to behave according to a multi-commodity isoelastic utility function $u_{\underline{R}}$, where $\underline{R} = (R_1,\ldots,R_d)$ is a risk-aversion vector with each risk-aversion value $R_k \in \R_{{\scriptscriptstyle \ge 0}}$ specific to lottery $k$ and satisfying either of the following conditions:
		\begin{equation}
			\label{eq:ra_condition2}
				\begin{aligned}
					\mathrm{(i)} &\ \mbox{$R_k\geq1$ 
						$(k=1, \ldots, d)$, }\\
					\mathrm{(ii)} &\ \mbox{$0< R_k< 1$ $(k=1, \ldots, d)$ and 
						$\sum\limits_{k=1}^{d}\!R_k>d-1$}.
				\end{aligned}
		\end{equation}
		With these elements in place, the economic-theoretic value that the gambler assigns to the $d$ lotteries in question is given by the isoelastic certainty equivalent (ICE) $w^{\rm ICE}_{\underline{R}} \equiv w^{\rm ICE}(p_{XG}^{(0)}, \vec O_X, \vec{B}_{X|G}, u_{\underline{R}})$ via \cref{eq:multiICE2}. 
		This ICE can be upper bounded as
		\begin{align}
			&\log
			w^{\rm ICE}(p_{XG}^{(0)}, \vec O_X, \vec{B}_{X|G}, u_{\underline{R}})\notag\\
			&
			\quad\le
			D_{\underline{\alpha},\alpha_0}(p_{X|G}^{(0)}, p_X^{(1)}, \ldots, p_X^{(d)}|p_G^{(0)})
			\nonumber
			\\
			&\qquad+
			\sum_{k=1}^{d}
			\frac{\alpha_k}{\alpha_0-1}
			\left(
			D_{S_{k}}
			\big(
			q_{XG}^{(k)}
			\big\|
			r^{(k)}_{XG}
			\big)
			+
			\log
			F^{(k)}
			\right)
			,
		\end{align}
		where $D_{\underline{\alpha},\alpha_0}(p_{X|G}^{(0)}, p_X^{(1)}, \ldots, p_X^{(d)}|p_G^{(0)})$ is the conditional multivariate R\'enyi divergence of PMFs $p_{X|G}^{(0)}$ and
		\begin{align*}
			p_X^{(k)}
			\coloneqq
			\frac{1}{
				F^{(k)}o_X^{(k)}
			},
			\ 
			F^{(k)}
			\coloneqq
			\sum_{x\in \mc X}
			\frac{1}{o_X^{(k)}(x)}\ \ \ (k=1, \ldots, d)
		\end{align*}
		the same as \cref{eq:E_ge1} conditioned by $p_G^{(0)}$, and $D_{S_k}(q_{XG}^{(k)}\|r_{XG}^{(k)})$ is the bivariate R\'enyi divergence \cref{eq:Renyi_bi} of joint PMFs 
		$q_{XG}^{(k)}=q_{G}^{(k)}q_{X|G}^{(k)}$ and $r_{XG}^{(k)}=q^{(k)}_{G}b_{X|G}^{(k)}$ with $q_{X|G}^{(k)}$ independent of $b_{X|G}^{(k)}$ (the explicit expressions are detailed in \cref{a:r2}).
		The orders $\underline{\alpha} = (\alpha_0, \ldots, \alpha_{d})$ and $S_k$ for these divergences are determined by the risk-aversion vector \cref{eq:ra_condition2} and given respectively as
		\begin{equation}
			\label{eq:ra_alpha2}
			\alpha_0
			\coloneqq\left(1+\sum_{k=1}^{d}(R_k-1)\right)^{-1}\!\!\!, \ \  \alpha_k\coloneqq(R_k-1) \alpha_0,
		\end{equation}
		and
		\begin{align}
			S_k
			\coloneqq 
			\frac{
				\sum_{l=0}^k \alpha_{l}
			}{
				\sum_{l=0}^{k-1} \alpha_{l}
			}\ \ (\ge0).
		\end{align}
	\end{result}
	
	This result can be proven using similar methods to \cref{r:r1}, and the full proof is presented in \cref{a:r2}.  
	Again, this statement holds very generally for all joint PMFs $p_{XG}^{(0)}$, all types of odds $\vec O_X$ proposed by the bookmakers, all types of betting strategies placed by the gambler and described by conditional PMFs $\vec{B}_{X|G}$, as well as for arbitrary agents with risk-averse attitudes described by the multi-commodity utility function $u_{\underline{R}}$ in \cref{eq:multi-IEU}. 
	Also, the order restriction is rephrased by the gambler's risk-averse attitude through \cref{eq:ra_condition2} and \cref{eq:ra_alpha2}.
	The major difference being that it is now the \emph{conditional} multivariate R\'{e}nyi divergence, the quantity that characterises the certainty equivalent. 
	Similarly to \cref{r:r1} we can now extract the optimal betting strategies in this generalised scenario.
	
	\begin{corollary}[\emph{Optimal betting strategies}] \label{c:c4}
		The log isoelastic certainty equivalent of a $d$-lottery betting game with side information and a gambler implementing optimal betting strategies reads
		\begin{align*}
			&\underset{\vec B_{X|G}}{\max}~
			\log
			w^{\rm ICE}(p_{XG}^{(0)}, \vec O_X, \vec{B}_{X|G}, u_{\underline{R}})
			\nonumber
			\\
			&\qquad=D_{\underline{\alpha},\alpha_0}  \big(
			\vec{P}_{X|G}
			\big|
			p_G^{(0)}
			\big)
			+
			\sum_{k=1}^{d}
			\frac{\alpha_k}{\alpha_0-1}
			\log
			F^{(k)},
		\end{align*}
		where the optimisation is taken over all betting strategies represented by conditional PMFs $\vec B_{X|G}$.
	\end{corollary}
	
	The analysis concerning the fairness of the odds is similar to that for \cref{r:r1} and we thus obtain the following corollary.
	\begin{corollary}[\emph{Conditional multivariate R\'enyi divergence characterises multi-lottery betting games, access to side information, optimal betting strategies, and fair odds}] \label{c:c5}
		The log isoelastic certainty equivalent of a $d$-lottery betting game with side information, fair odds, and a gambler implementing optimal betting strategies
		reads
		\begin{equation*}
			\begin{aligned}
				&\underset{\vec B_{X|G}}{\max}~
				\log
				w^{\rm ICE}(p_{XG}^{(0)}, \vec O_X^{\mathrm{Fair}}, \vec{B}_{X|G}, u_{\underline{R}})\\
				&\qquad\qquad\qquad=D_{\underline{\alpha},\alpha_0}
				\big(
				\vec{P}_{X|G}
				\big|
				p_G^{(0)}
				\big),
			\end{aligned}
		\end{equation*}
		where the optimisation is taken over all betting strategies.
	\end{corollary}
	
	This provides an interpretation of the \emph{conditional} multivariate R\'{e}nyi divergence as the economic-theoretic value that a rational agent assigns to a multi-lottery betting game, when all the bookmakers propose fair odds, when the agent has access to side information, and when the agent is allowed to maximise over betting strategies. 
	Taking into account that there can be multiple ways of defining conditional multivariate R\'enyi divergences, one technical aspect worth highlighting is that this approach singles out a specific conditional multivariate R\'enyi divergence  (the BLP type) and, amongst these, it furthermore specifies the second order to be $\alpha_0 = \max_{k} \alpha_k$. 
	Finally, we can now also provide an economic-theoretic interpretation of the data processing inequality (\cref{c:c1}) as follows.
	\begin{corollary}[\emph{Economic-theoretic interpretation of the data processing inequality}] \label{c:c6}
		We compare a scenario of a $d$-lottery betting game where the gambler has access to some side information $G$ 
		against a situation where the gambler does not have access to side information. In both situations, the gambler is allowed to maximise over all betting strategies (over conditional PMFs $\vec B_{X|G}$ when having access to $G$, and over PMFs $B_X$ when there is no side information). 
		Using the data processing inequality for the conditional multivariate R\'enyi divergence, we have for their log isoelastic certainty equivalents
		\begin{multline}
			\underset{\vec B_{X|G}}{\max}~
			\log
			w^{\rm ICE}(p_{XG}^{(0)}, \vec O_X^{\mathrm{Fair}}, \vec{B}_{X|G}, u_{\underline{R}})
			\\
			\geq
			\underset{\vec B_{X}}{\max}~
			\log
			w^{\rm ICE}(p_{X}^{(0)}, \vec O_X^{\mathrm{Fair}}, \vec{B}_{X}, u_{\underline{R}})
			.
		\end{multline}
		In other words, having access to side information increases the economic-theoretic value that the gambler assigns to the $d$ lotteries.
	\end{corollary}
	We therefore observe that having access to side information (\cref{r:r2}) increases the certainty equivalent when compared to having no side information (\cref{r:r1}). In words, side information increases the ``value" of the $d$ lotteries in question for a risk-averse gambler.
	We again emphasise that the gambler's economic-theoretically consistent attitude \cref{eq:ra_condition2} of risk aversion for every lottery now implies the data processing inequality as an information-theoretic consequence.
	Specifically, \cref{c:c5} and \cref{c:c6} reveal a quantitative connection between  the economic-theoretic benefit and the data processing inequality through the \emph{conditional} multivariate R\'enyi divergence.
	This finishes our presentation of the results concerning the characterisation of multi-lottery betting games in terms of multivariate unconditional and conditional R\'enyi divergences. 
	We are now ready to show the applicability of these results within the operational framework of general probabilistic theories and reveal the connection between the information, economic, and physical theories.
	\section{State betting games in general probabilistic theories and multivariate R\'enyi divergences}
	\label{s:gpt1}
	
	
	\subsection{General probabilistic theories (GPTs)}
	We have so far considered betting on classical labels.
	In this section, we introduce the task of \textit{state betting}, where the gambler bets on
	states being allowed to perform measurements as side information for the bets.
	This scenario is well-described by notions in \textit{general probabilistic theories (GPTs)}.
	A \textit{GPT} is a pair of subsets $(\Omega,\mathcal{E})$ of a finite-dimensional real Euclidean space $V$.
	The set $\Omega$, called a \textit{state space}, is a compact and convex subset of $V$ whose linear and affine hulls satisfy $\mathit{lin}(\Omega)=V$ and $\mathit{aff}(\Omega)\not\owns O$ (the origin of $V$) respectively.
	Elements of $\Omega$ are called \textit{states} and they mathematically represent physical systems.
	On the other hand, the set $\mathcal{E}$, called an \textit{effect space}, is defined as $\EE=\{m\in V\mid\ang{m,\omega}\in[0,1],\forall\omega\in\Omega\}$ and its elements are called \textit{effects}.
	Here $\ang{m,\omega}$ denotes the Euclidean inner product of two vectors $\omega,m\in V$.
	The element $u$ of $\mathcal{E}$ satisfying $\ang{u,\omega}=1~(\forall\omega\in\Omega)$ is called the \textit{unit effect}.
	A \textit{measurement} $\M$ with an outcome set $\mc A$ is defined as a family of effects (more precisely, an \textit{effect-valued measure}) $\M=\{m_a\}_{a\in \mc A}$ such that $\sum_{a\in \mc A}m_a=u$.
	The measurement $\M$ is a mathematical representation of a measurement apparatus.
	The output $\{\ang{m_a,\omega}\}_{a\in \mc A}$ represents the statistics observed when the measurement is performed using the apparatus on a physical system described by a state $\omega\in\Omega$.
	In particular, each effect $m_a$ outputs the probability of observing an outcome $a\in \mc A$ in the experiment.
	Throughout this article, we assume the \textit{no-restriction hypothesis} \cite{Chiribella2010} for mathematical simplicity; we assume that all elements of $\EE$ are physically valid, but this requirement is not imposed in certain settings \cite{Janotta2013,Sainz2018,Filippov2020}.

	The most important and physically relevant examples of GPTs are quantum and classical probability theories.
	With an $n$-dimensional complex Hilbert space $\mathscr{H}$ ($n<\infty$), a finite-dimensional quantum theory is described in terms of the $n^2$-dimensional real vector space $\mathcal{L}_S(\mathscr{H})$ of self-adjoint operators on $\mathscr{H}$. 
	The space $\mathcal{L}_S(\mathscr{H})$ is regarded as a Euclidean space equipped with the Hilbert-Schmidt inner product $\ang{X,Y}_{\mathrm{HS}}=\Tr[X^\dagger Y]$.
	A finite-dimensional quantum theory is then given by the pair $(\state(\mathscr{H}), \mathcal{E}(\mathscr{H}))$, where $\state(\mathscr{H})$ is the set of all positive semi-definite and unit-trace operators (density operators) on $\mathscr{H}$ and $\mathcal{E}(\mathscr{H})=\{E\in\mathcal{L}_S(\mathscr{H})\mid\ang{\rho,E}_{\mathrm{HS}}\in[0,1],~\forall\rho\in\state(\mathscr{H})\}$.
	We note that $E\in \mathcal{E}(\mathscr{H})$ if and only if $E$ is bounded by the zero operator $O$ and the identity operator (the unit effect) $I$ on $\mathscr{H}$, namely $O\le E\le I$ \cite{Heinosaari2011,Busch2016}.
	Measurements in the theory $(\state(\mathscr{H}), \mathcal{E}(\mathscr{H}))$ are called positive operator-valued measures (POVMs).
	On the other hand, a finite-dimensional classical probability theory is described by a probability simplex \cite{Boyd2004}.
	With $V=\R^n$, it is given as $(\mathcal{P}_n,\mathcal{E}(\mathcal{P}_n))$, where $\mathcal{P}_n=\{(p_i)_{i=1}^n\in V\mid p_i\ge0,~ \sum_ip_i=1 \}$ is the set of all $n$-outcome PMFs and $\mathcal{E}(\mathcal{P}_n)=\{(v_i)_{i=1}^n\in V\mid v_i\in[0,1] \}$.
	This finishes a concise summary of GPTs. A more rigorous formulation and advanced topics can be found in \cite{Plavala2023,Lami2017,Takakura2022}.

	\subsection{Economic-theoretic values of multi-lottery state betting games}
	Let us move on to generalising betting games discussed in previous sections and introduce \textit{state betting games}.
	The task of state betting was first proposed in the realm of quantum theory \cite{QBG_2022}.
	In this article, we generalise the scenario in terms of multiple lotteries and states in GPTs.
    \begin{figure}
		\includegraphics[scale=1.56]{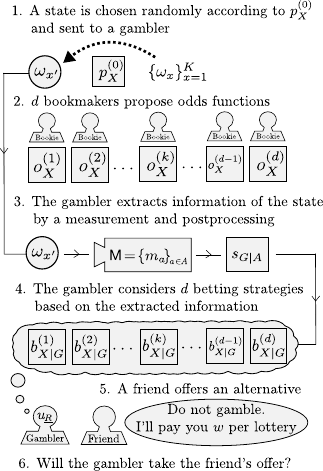}
		\caption{
			Illustration of a multi-lottery state betting game. 
            Instead of a classical horse race, a state is randomly chosen according to an ensemble of states $\Lambda=\{p_X^{(0)}(x),\omega_x\}_x$. 
            A group of $d$ bookmakers propose $d$ odds functions $\vec O_X=(o_X^{(1)},\ldots,o_X^{(d)})$, and the gambler extracts information of the state from a measurement $\mathsf{M}=\{m_a\}_a$ and a postprocessing strategy $s_{G|A}$ to place $d$ betting strategies $\vec{B}_{X|G}=(b_{X|G}^{(1)},\ldots,b_{X|G}^{(d)})$.
			The economic-theoretic value that the gambler assigns to the gambling is given by the isoelastic certainty equivalent
            $w^{\rm ICE}_{}(\M,\Lambda,s_{G|A},\vec{B}_{X|G},\vec{O}_X,u_{\vec R})$.
		}
		\label{fig:SB}
	\end{figure}

	Consider again three types of agents: a referee, a gambler, and $d$ bookmakers. 
	This time, instead of classical labels, the referee has an ensemble $\Lambda$ of states of a GPT $(\Omega,\EE)$, namely, $\Lambda=\{p_X^{(0)}(x),\omega_x\}_{x\in \mc X}$ with $\omega_x\in\Omega$ and $p_X^{(0)}$ a PMF on $\mc X$.
	The referee starts the game by sending the gambler one of the states, say $\omega_{x'}$, according to $p_X^{(0)}$. 
	The goal of the game is for the gambler to correctly identify the state (or the index $x'$) that was sent.
	In order to do this, the gambler is allowed to perform a measurement $\M=\{m_a\}_{a\in \mc A}$ and extract information from the state. 
	The gambler is also allowed to postprocess the measurement outcome $a$ to produce an index $g$ as the guess of the state. 
	Until this point, the game is essentially the same as standard state discrimination (SD), where the gambler ``wins" whenever $g=x'$ and ``loses" otherwise. 
	In SD, the quantity of interest is the probability of successfully identifying the correct state.
	It is calculated as
	\begin{equation*}
		p^{\rm SD}_{\rm succ}
		(\M,\Lambda)
		=
		\max_{s_{G|A}}
		\sum_{g,a,x}
		\delta_{x,g}\,
		s_{G|A}(g|a)
		\ang{m_a, \omega_x}p_X^{(0)}(x)
		,
	\end{equation*}
	where the optimisation is over all postprocessing strategies $s_{G|A}$ represented by conditional PMFs. 
	We now generalise the task of SD incorporating the idea of betting games. 
	In a \textit{state betting (SB) game}, the gambler does not provide a unique index $g$, as a guess of the state that was sent, but instead to ``bet" on all possible outcomes according to the measurement and its postprocessing. 
	The scenario closely resembles the betting games discussed so far.
	Here we further employ the multi-commodity setting to constitute a \textit{multi-lottery state betting game}.
	That is, $d$ bookmakers propose $d$ odds functions $\vec{O}_X=(o_X^{(1)},\ldots,o_X^{(d)})$ in addition to the referee's ensemble $\Lambda=\{p_X^{(0)}(x),\omega_x\}_{x\in \mc X}$.
	On the other hand, the gambler performs the measurement $\M=\{m_a\}_{a\in \mc A}$ on the received state $\omega_{x'}$ and postprocesses the outcome through the conditional PMF $s_{G|A}$ to obtain a label $g\in \mc G$.
	The gambler then provides $d$ betting strategies $\vec{B}_{X|G=g}=(b_{X|G=g}^{(1)},\ldots,b_{X|G=g}^{(d)})$ in the form of conditional PMFs $b_{X|G=g}^{(k)}=\{b_{X|G}^{(k)}(x|g)\}_{x\in \mc X}$ on $\mc X$ according to the side information $g$ on the state.
	After placing the $d$ bets $\vec{B}_{X|G=g}$, the referee reveals the state that was sent (or the index $x'$), and the bookmakers reward the gambler with multi-commodity wealth $(w^{(1)}_{XG}(x',g),\ldots,w^{(d)}_{XG}(x',g))$, where
	$w^{(k)}_{XG}(x', g)=b^{(k)}_{X|G=g}(x'|g)o^{(k)}_{X}(x')$.
	This SB scenario can be interpreted as a $d$-lottery betting game with side information, where the side information is conveyed by the state and extracted by the gambler by means of the measurement and postprocessing such that $p_{XG}^{(0)}[\M,\Lambda,s_{G|A}](x,g) = \sum_{a\in \mc A}s_{G|A}(g|a)\ang{m_a, \omega_x} p_X^{(0)}(x)$.

	Now let us present another main result on the economic-theoretic ``value" of the multi-lottery SB game.
	Our analysis above implies that the SB game can directly be related to the variant of a $d$-lottery betting game which involves the agent having access to side information. 
	In particular, the scenario is identified with the gambler playing $d$ lotteries $ \{(p_{XG}^{(0)}[\M,\Lambda,s_{G|A}], b_{X|G}^{(k)}o_X^{(k)})\}_{k=1}^d\equiv (\M,\Lambda,s_{G|A},\vec O_X,\vec{B}_{X|G})$, 
	and the value of the lotteries is given by the isoelastic certainty equivalent $w^{\rm ICE}_{}(\M,\Lambda,s_{G|A},\vec{B}_{X|G},\vec{O}_X,u_{\vec R})$, which is defined via \cref{eq:multiICE2} with the multi-commodity isoelastic utility function $u_{\vec R}$ in \cref{eq:multi-IEU}.
	This motivates us to rephrase \cref{r:r2} in terms of SB.
	\begin{result}[\emph{Economic-theoretic characterisation of the multi-lottery state betting game through the multivariate R\'enyi divergence}]
		\label{r:r3}
		Consider a GPT $(\Omega,\EE)$ and a $d$-lottery state betting (SB) game involving a referee, $d$ bookmakers ($d\ge1$), and a gambler.
		The referee possesses an ensemble of states $\Lambda = \{p_X^{(0)}(x),\omega_x\}_{x\in \mc X}$ to choose one of the states according to the PMF $p_X^{(0)}$ and send it to the gambler.
		The $d$ bookmakers offer $d$ odds functions $\vec{O}_X=(o_X^{(1)},\ldots,o_X^{(d)})$.
		The gambler is allowed to perform a measurement $\M = \{m_a\}_{a\in \mc A}$ on the received state and its postprocessing via $s_{G|A}$, and use them to place $d$ betting strategies $\vec{B}_{X|G}=(b_{X|G}^{(k)})_{k=1}^{d}$.
		It effectively establishes $d$ lotteries $ \{(p_{XG}^{(0)}[\M,\Lambda,s_{G|A}], b_{X|G}^{(k)}o_X^{(k)})\}_{k=1}^d\equiv (\M,\Lambda,s_{G|A},\vec O_X,\vec{B}_{X|G})$ with $p_{XG}^{(0)}[\M,\Lambda,s_{G|A}](x,g) = \sum_{a\in \mc A}s_{G|A}(g|a)\ang{m_a, \omega_x} p_X^{(0)}(x) $.
		The gambler is also assumed to behave according to a multi-commodity isoelastic utility function $u_{\underline{R}}$, where $\underline{R} = (R_1,\ldots,R_d)$ is a risk-aversion vector with each risk-aversion value $R_k \in \R_{{\scriptscriptstyle \ge 0}}$ specific to lottery $k$ and satisfying either of the following conditions:
		\begin{equation}
			\label{eq:ra_condition3}
				\begin{aligned}
					\mathrm{(i)} &\ \mbox{$R_k\geq1$ 
						$(k=1, \ldots, d)$, }\\
					\mathrm{(ii)} &\ \mbox{$0< R_k< 1$ $(k=1, \ldots, d)$ and 
						$\sum\limits_{k=1}^{d}\!R_k>d-1$}.
				\end{aligned}
		\end{equation}
		With these elements in place, the economic-theoretic value that the gambler assigns to the $d$ lotteries (equivalently, the SB game) is given by the isoelastic certainty equivalent (ICE) $w^{\rm ICE}_{\underline{R}} \equiv w^{\rm ICE}(\M,\Lambda,s_{G|A}, \vec O_X, \vec{B}_{X|G}, u_{\underline{R}})$. 
		This ICE can be upper bounded as
		\begin{align}
			&\log w^{\rm ICE}(\M,\Lambda,s_{G|A}, \vec O_X, \vec{B}_{X|G}, u_{\underline{R}})\notag\\
			&\leq 
			D_{\underline{\alpha},\alpha_0}  \big(
			p^{(0)}_{X|G}[\M,\Lambda,s_{G|A}], p_X^{(1)}, \ldots,p^{(d)}_{X}
			\big|
			p_G^{(0)}[\M,\Lambda,s_{G|A}]\big)\notag\\
			&\quad
			+
			\sum_{k=1}^{d}
			\frac{\alpha_k}{\alpha_0-1}
			\left(
			D_{S_{k}}
			\big(
			q_{XG}^{(k)}
			\big\|
			r_{XG}^{(k)}
			\big)
			+
			\log{F^{(k)}}
			\right),\label{eq:result3}
		\end{align}
		where the PMFs
		$p_{X|G}^{(0)}[\M,\Lambda,s_{G|A}]$ and $p_{G}^{(0)}[\M,\Lambda,s_{G|A}]$ are determined through the joint PMF $p_{XG}^{(0)}[\M,\Lambda,s_{G|A}]$ and
		\begin{align*}
			p_X^{(k)}
			\coloneqq
			\frac{1}{
				F^{(k)}o_X^{(k)}
			},
			\ 
			F^{(k)}
			\coloneqq
			\sum_{x\in \mc X}
			\frac{1}{o_X^{(k)}(x)}\ \ \ (k=1, \ldots, d)
		\end{align*}
		the same as \cref{eq:E_ge1}, and the joint PMFs $q_{XG}^{(k)}$ and $r_{XG}^{(k)}$ are of the form
		$q_{XG}^{(k)}=q_{G}^{(k)}q_{X|G}^{(k)}$ and $r_{XG}^{(k)}=q^{(k)}_{G}b_{X|G}^{(k)}$ with $q_{X|G}^{(k)}$ independent of $b_{X|G}^{(k)}$ (the explicit expressions are detailed in \cref{a:r2}).
		The orders $\underline{\alpha} = (\alpha_0, \ldots, \alpha_{d})$ and $S_k$ for the divergences $D_{\underline{\alpha},\alpha_0}$ and $D_{S_k}$ are determined by the risk-aversion vector \cref{eq:ra_condition3} and given respectively as
		\begin{equation*}
			\alpha_0
			\coloneqq\left(1+\sum_{k=1}^{d}(R_k-1)\right)^{-1}\!\!\!, \ \  \alpha_k\coloneqq(R_k-1) \alpha_0,
		\end{equation*}
		and
		\begin{align}
			S_k
			\coloneqq 
			\frac{
				\sum_{l=0}^k \alpha_{l}
			}{
				\sum_{l=0}^{k-1} \alpha_{l}
			}\ \ (\ge0).
		\end{align}
	\end{result}
	We omit the proof because it proceeds in the same way as \cref{r:r2}.
	We note that the PMF $p_{XG}^{(0)}[\M,\Lambda,s_{G|A}]\colon (x,g)\mapsto \sum_{a}s_{G|A}(g|a)\ang{m_a, \omega_x} p_X^{(0)}(x)$ on $\mc X\times \mc G$ is postprocessing of another PMF $p_{XA}^{(0)}[\M,\Lambda]\colon (x,a)\mapsto\ang{m_a, \omega_x} p_X^{(0)}(x)$ on $\mc X\times \mc A$ via $s_{G|A}$.
	It enables us to apply the data processing inequality in \cref{l:l2} and optimise the ICE on all postprocessing strategies.
	\begin{corollary}[\emph{Optimal betting and postprocessing strategies for the state betting game}]
		\label{cor:r3}
		When optimising over the betting $\vec{B}_{X|G}$ and postprocessing strategies $s_{G|A}$, the upper bound of \cref{eq:result3} can be reached:
		\begin{align*}
			&\underset{\vec B_{X|G},\,s_{G|A}}{\max}
			\log w^{\rm ICE}(\M,\Lambda,s_{G|A}, \vec O_X, \vec{B}_{X|G}, u_{\underline{R}})
			\nonumber
			\\
			&\quad=
			D_{\underline{\alpha},\alpha_0}  \big(
			p^{(0)}_{X|A}[\M,\Lambda], p_X^{(1)}, \ldots,p^{(d)}_{X}
			\big|
			p_A^{(0)}[\M,\Lambda]\big)\notag\\
			&\qquad
			+
			\sum_{k=1}^{d}
			\frac{\alpha_k}{\alpha_0-1}
			\log{F^{(k)}},
		\end{align*}
		where $p_{X|A}^{(0)}[\M,\Lambda]$ and $p_{A}^{(0)}[\M,\Lambda]$ are determined through the joint PMF
		$p_{XA}^{(0)}[\M,\Lambda](x,a)=p_X^{(0)}(x)\ang{m_a,\omega_x}$.
		In particular, when the odds are fair ($F^{(k)} = 1$, $\forall k$), it reduces to
		\begin{align*}
			&\underset{\vec B_{X|G},\,s_{G|A}}{\max}\log{w^{\rm ICE}}
			(\M,\Lambda,s_{G|A}, \vec O_X^{\mathrm{Fair}}, \vec{B}_{X|G}, u_{\underline{R}})
			\\
			&\qquad=D_{\underline{\alpha},\alpha_0}  \big(
			p^{(0)}_{X|A}[\M,\Lambda], p_X^{(1)}, \ldots,p^{(d)}_{X}
			\big|
			p_A^{(0)}[\M,\Lambda]\big)
		\end{align*}
		with $p^{(k)}_X = (o_X^{(k)})^{-1}$ ($k=1, \ldots, d$).
	\end{corollary}
	
	The results again provide a practical interpretation of the multivariate R\'enyi divergence through the economic-theoretic value of a task involving betting with multiple lotteries.
	What is specific to the current situation is that the influence of states (precisely, an ensemble) $\Lambda$ and a measurement $\M$ on the task is incorporated through the PMFs $p_{XG}^{(0)}[\M,\Lambda,s_{G|A}]$ and $p_{XA}^{(0)}[\M,\Lambda]$.
	In other words, here we have revealed a quantitative connection between information-theoretic, economic-theoretic, and physical notions.
	In the subsequent section, we will develop a resource-theoretic aspect of the task of SB.

	Let us conclude the section with several remarks.
	First, we can check that SB games generalise SD. 
	In fact, if we restrict the single-lottery scenario ($d=1$), set the odds as constant $o_X^{(1)}(x)\coloneqq C>0$ ($\forall x$), and the gambler to have a neutral attitude towards risk ($R=0$), it is straightforward to check that the optimal CE becomes
	\begin{equation}
		\label{eq:ICE-SD}
		\max_{b_{X|G}^{(1)}, \,s_{G|A}}\!\!\!\!
		w^{\rm ICE}_{R=0}(\M,\Lambda,s_{G|A},C,b^{(1)}_{X|G})
		=
		C
		p^{\rm SD}_{\rm succ}
		(\M,\Lambda),
	\end{equation}
	where $b^{(1)}_{X|G}(x|g)=\delta_{x,g}$ realises the maximum over all $b^{(1)}_{X|G}$.
	Second, the SB discussed here is a natural extension of \textit{quantum state betting (QSB)} introduced in \cite{QBG_2022} to GPTs.
	When the GPT is quantum, namely $(\Omega,\EE)=(\state(\mathscr{H}),\EE(\mathscr{H}))$, the task of SB reduces to QSB.
	Also, (multi-lottery) betting games are reproduced simply by applying a classical probability theory $(\Omega,\EE)=(\mathcal{P}_n,\mathcal{E}(\mathcal{P}_n))$.
	These observations reinforce the operational aspect of the multivariate R\'enyi divergence.

	\section{Resource-theoretic expressions of state betting games}
	\label{s:gpt2}
\begin{figure}
		\begin{center}
			\begin{overpic}[scale=0.35,unit=1mm]{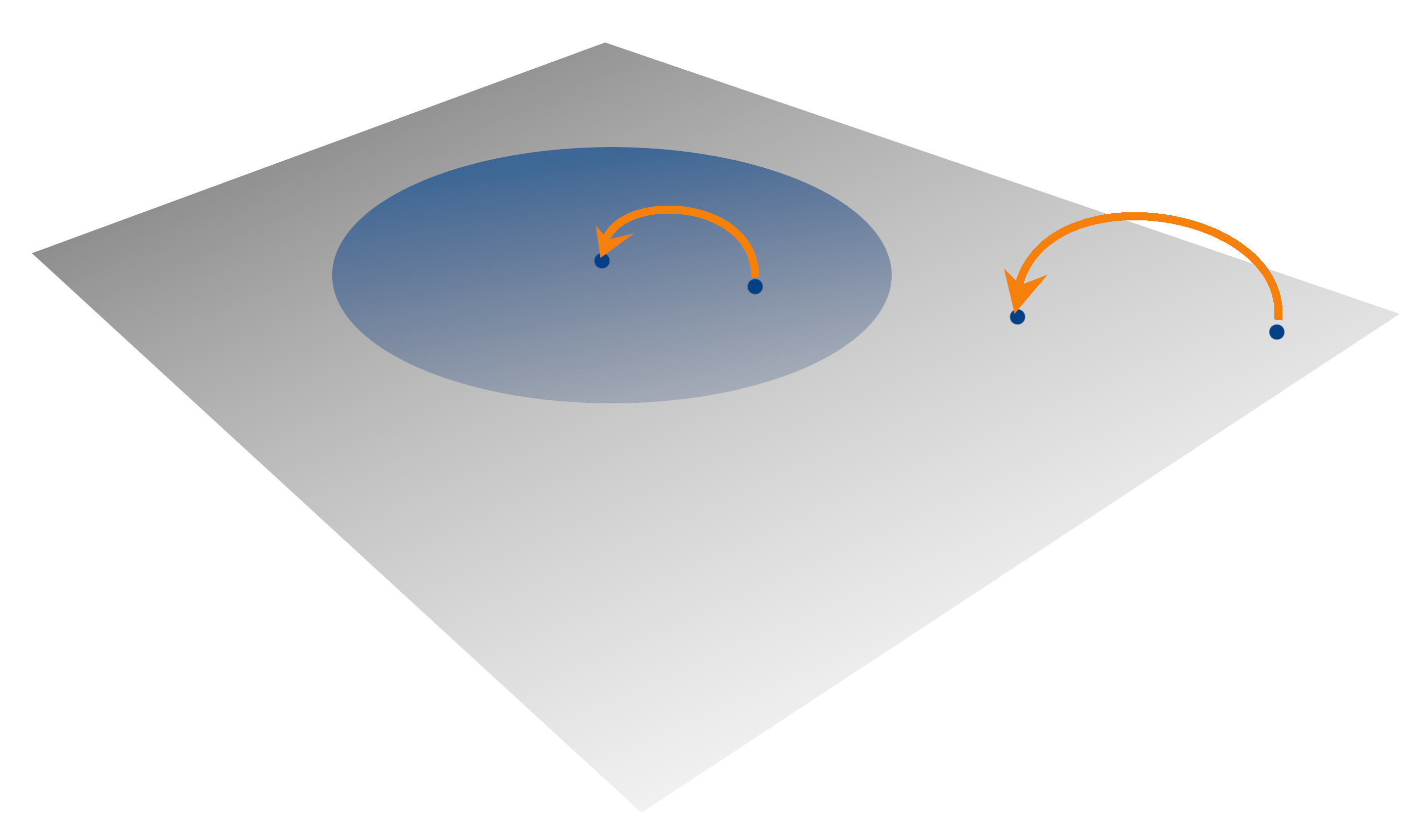}
				\put(73,25){\begin{large}
						$x$
				\end{large}}
				\put(57,26){\begin{large}
						$x'$
				\end{large}}
				\put(65,38){\begin{large}
						$O$
				\end{large}}
				\put(42,28){\begin{large}
						$y$
				\end{large}}
				\put(32,30){\begin{large}
						$y'$
				\end{large}}
				\put(39,39){\begin{large}
						$O$
				\end{large}}
				\put(35,15){\begin{Huge}
						$\mc R$
				\end{Huge}}
				\put(20,30){\begin{Huge}
						$\mc F$
				\end{Huge}}
			\end{overpic}
		\end{center}
		\caption{\label{fig:resource} 
			A depiction of a resource theory with resource set $\mc R$ and free set $\mc F$. A free operation $O\in\mc O_{\mc F}$ is required to map a free resource $y\in\mc F$ into a free resource. 
			A genuine resource $x\in\mc R\setminus\mc F$ is mapped by $O$ often in some sense ``closer'' to the free set, and this kind of monotonicity is quantitatively expressed by the notion of resource monotones.}
	\end{figure}
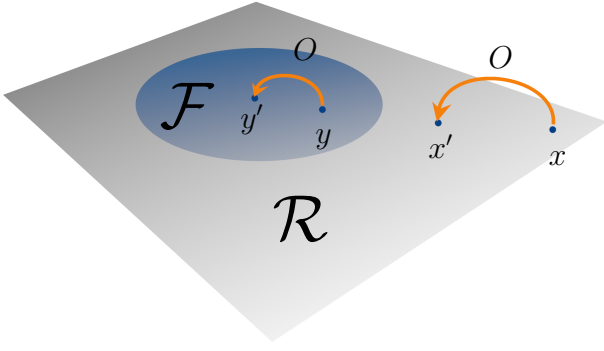

	Let us now analyse multi-lottery SB games from the perspective of \textit{resource theories (RTs)}. 
	In this paper, we mainly study RTs in the realm of GPTs  generalising quantum resource theories.  
	The basis of any RT is the set $\mc R$ of all resources. 
	This set may consist of various elements such as states and measurements. 
	Mathematically, $\mc R$ is often modelled as a subset of an affine plane or a vector space; 
	For states and measurements, $\mc R$ is a subset of the underlying vector space that embeds the state and effect spaces respectively.
	Among the resources in $\mathcal{R}$, some are considered to possess no resource.
	The set of all such elements $\mathcal{F} \subset \mathcal{R}$ is called the \textit{free set}, because we usually think of zero-resource elements as available to us without restriction. 
	In contrast, elements of $\mathcal{R}\setminus \mc F$ are called \textit{resourceful}.
	The free set $\mc F$ is often convex, in which case the RT is also called a {\it convex resource theory}. 
	Another important concept in RTs is the set $\mc O_{\mc F}$ of {\it free operations}, suitable maps on $\mc R$ which are somehow freely available. 
	Essentially always a minimal requirement for $O\in\mc O_{\mc F}$ is that $O(x)\in\mc F$ whenever $x\in\mc F$, i.e., free maps cannot create resources out of free elements.  
	\cref{fig:resource} gives a very general picture of RTs.
	The main object of this section is the RT of informative measurements \cite{PS_NL_2019, RT_BS_2019}.
	\begin{definition}[\emph{Resource theory of informative measurements}]
		\label{def:RT of info}
		Consider a GPT $(\Omega,\EE)$. 
		The set $\mc R$ of all measurements $\M=\{m_a\}_{a\in\mc A}$ on a discrete outcome set constitutes the set of all resources. 
		We say that a measurement $\NN=\{n_a\}_{a\in \mc A}\in\mc R$ is {\it uninformative} (or trivial) if all effects $n_a$ are multiples of the unit effect $u$, i.e., $n_a=q_A(a)u$ for all $a\in\mc A$ with a PMF $q_A\colon \mc A\to[0,1]$. 
		The resource theory with its free set $\mathbb{UI}\subset\mc R$ of uninformative measurements is called the {\it resource theory of informative measurements}. 
		As a set $\mc O_{\mathbb{UI}}$ of free operations, we consider postprocessing maps introduced as follows:
		We pick another discrete set $\mc B$ and a conditional PMF $t_{B|A}$.
		We define a map $T_{B|A}$ that maps a measurement $\M=\{m_a\}_{a\in\mc A}$ with outcomes $\mc A$ to a measurement $T_{B|A}(\M)=\{m'_b\}_{b\in\mc B}$ with outcomes $\mc B$, where
		\begin{equation}
			\label{eq:free op}
			m'_b=\sum_{a\in \mc A}t_{B|A}(b|a)m_a.
		\end{equation}
		These postprocessing maps clearly satisfy $T_{B|A}(\NN)\in\mathbb{UI}$ whenever $\NN\in\mathbb{UI}$.
	\end{definition}
	
	A central question in RTs is to find reasonable quantifications of resources.
	Such quantifiers are often required to be monotonic with respect to free operations.
	More precisely, for a RT $(\mc R, \mc F, \mc O_{\mc F})$, a function $f\colon \mc R\to\R$ on the resource set $\mc R$ is called a \emph{resource measure} (or \emph{resource monotone}) if $f(x)\ge 0$ for any $x\in\mc R$, $f(x)=0$ whenever $x\in \mc F$, and $f(x)\ge f(O(x))$ for any $x\in\mc R$ and $O\in\mc O_{\mc F}$ \cite{Chitambar2019,Gour2025}.
	Among resource measures, information-theoretically and physically (operationally) meaningful quantifiers are particularly important. 
	Here we propose a resource measure for informative measurements based on multivariate R\'enyi divergence and reveal its practical interpretation through the task of multi-lottery SB games.
	\begin{widetext}
		\begin{result}[\emph{Resource measures for informative measurements based on multi-lottery state betting games}]
			\label{r:r4}
			Let $\underline{\alpha}=(\alpha_0,\ldots,\alpha_{d})\in\R^{d+1}$ be orders satisfying $\alpha_0=\max_{0\le k\le d}\alpha_k$, $\sum_{k=0}^d\alpha_k=1$, and one of the conditions of \cref{eq:parameter}.
			Consider a GPT $(\Omega,\mc E)$. 
			For an ensemble of states $\Lambda=\{p_X^{(0)},\omega_x\}_{x\in\mc X}$ and a measurement $\M=\{m_a\}_{a\in\mc A}$, we define a joint PMF $p_{XA}^{(0)}[\M,\Lambda](x,a) = \sum_{a\in \mc A}\ang{m_a, \omega_x} p_X^{(0)}(x) $ on $\mc X\times\mc A$.
			Then, for any uninformative measurement	$\NN=\{n_b\}_{b\in \mc B}\in\mathbb{UI}$, ensemble of states $\Lambda$, and $d$ PMFs  $\{r_X^{(k)}\}_{k=1}^d$ on $\mc X$, the multivariate R\'enyi divergences satisfy
			\begin{equation*}
				D_{\underline{\alpha},\alpha_0}  \big(
				p^{(0)}_{X|B}[\NN,\Lambda], r_X^{(1)}, \ldots,r^{(d)}_{X}
				\big|
				p_B^{(0)}[\NN,\Lambda]\big)=
				D_{\underline{\alpha}}  \big(
				p^{(0)}_X, r_X^{(1)}, \ldots,r^{(d)}_{X}
				\big),
			\end{equation*}
			and thus the quantity 
			\begin{equation}
				\mc D_{\underline{\alpha},\alpha_0}(\M,\Lambda,\{r_X^{(k)}\}_{k=1}^d):=D_{\underline{\alpha},\alpha_0}  \big(
				p^{(0)}_{X|B}[\M,\Lambda], r_X^{(1)}, \ldots,r^{(d)}_{X}
				\big|
				p_B^{(0)}[\M,\Lambda]\big)-
				D_{\underline{\alpha}}  \big(
				p^{(0)}_X, r_X^{(1)}, \ldots,r^{(d)}_{X}
				\big)
			\end{equation}
			is a resource measure for the resource theory of informative measurements.
			That is, $\mc D_{\underline{\alpha},\alpha_0}(\M,\Lambda,\{r_X^{(k)}\})\ge0$ and $D_{\underline{\alpha},\alpha_0}(\M,\Lambda,\{r_X^{(k)}\})=0$ whenever $\M\in\mathbb{UI}$, and 
			for any free operation $T_{B|A}$ (see \cref{eq:free op}), the monotonicity
			\begin{equation}
				\label{eq:RT_monotone}
				\mc D_{\underline{\alpha},\alpha_0}(\M,\Lambda,\{r_X^{(k)}\})\ge \mc D_{\underline{\alpha},\alpha_0}(T_{B|A}(\M),\Lambda,\{r_X^{(k)}\})
			\end{equation}
			holds.
			The measure $\mc D_{\underline{\alpha},\alpha_0}$ can be practically interpreted as follows.
			Consider a $d$-lottery state betting (SB) game involving a referee, $d$ bookmakers ($d\ge1$), and a gambler.
			The referee possesses an ensemble of states $\Lambda = \{p_X^{(0)}(x),\omega_x\}_{x\in \mc X}$ to choose one of the states according to the PMF $p_X^{(0)}$ and send it to the gambler.
			The $d$ bookmakers offer $d$ odds functions $\vec{O}_X=(o_X^{(1)},\ldots,o_X^{(d)})$.
			The gambler is allowed to perform a measurement $\M = \{m_a\}_{a\in \mc A}$ on the received state and its postprocessing via $s_{G|A}$, and use them to place $d$ betting strategies  $\vec{B}_{X|G}=(b_{X|G}^{(k)})_{k=1}^{d}$.
			It effectively establishes $d$ lotteries $ \{(p_{XG}^{(0)}[\M,\Lambda,s_{G|A}], b_{X|G}^{(k)}o_X^{(k)})\}_{k=1}^d\equiv (\M,\Lambda,s_{G|A},\vec O_X,\vec{B}_{X|G})$ with $p_{XG}^{(0)}[\M,\Lambda,s_{G|A}](x,g) = \sum_{a\in \mc A}s_{G|A}(g|a)\ang{m_a, \omega_x} p_X^{(0)}(x) $.
			The gambler is assumed to behave according to a multi-commodity isoelastic utility function $u_{\underline{R}}$ with a risk-aversion vector $\underline{R} = (R_1,\ldots,R_d)\in\R^d_{{\scriptscriptstyle \ge0}}$ defined as
			\begin{equation*}
				R_k=1+\frac{\alpha_{k}}{\alpha_0}\qquad (k=1,\ldots,d).
			\end{equation*} 
			We consider the ratio
			\begin{equation}
				\mathcal{W}_{\mathbb{UI}}(\M,\Lambda,\vec{O}_X,u_{\underline{R}})			
				:=\frac{\displaystyle
					\max_{
						\vec{B}_{X|G},\,s_{G|A}
					}w^{\rm ICE}_{}\big(\M,\Lambda,s_{G|A},\vec{O}_X,\vec{B}_{X|G},u_{\underline{ R}}\big)
				}
				{
					\displaystyle
					\max_{\NN\in\mathbb{UI}}~
					\max_{
						\vec{B}_{X|G},\,s_{G|B}
					}
					\!\!\!w^{\rm ICE}_{}\big(\NN,\Lambda,s_{G|B},\vec{O}_X,\vec{B}_{X|G},u_{\underline{ R}}\big)
				}
			\end{equation}
			of the optimal isoelastic certainty equivalent representing the figure of merit in the SB game provided by a measurement $\M$ with outcomes $\mc A$ against all free measurements $\NN$ with outcomes $\mc B$.
			Then, with PMFs
			\begin{align*}
				p_X^{(k)}
				\coloneqq
				\frac{1}{
					F^{(k)}o_X^{(k)}
				},
				\quad
				F^{(k)}
				\coloneqq
				\sum_{x\in \mc X}
				\frac{1}{o_X^{(k)}(x)}
				\qquad
				(k=1,\ldots,d)
			\end{align*}
			determined by the odds $\vec{O}_X$, the resource measure $\mc D_{\underline{\alpha},\alpha_0}(\M,\Lambda,\{p_X^{(k)}\}_{k=1}^d)\equiv\mc D_{\underline{\alpha},\alpha_0}(\M,\Lambda,\vec{O}_X)$ is expressed as
			\begin{equation}
				\label{eq:ratio3}
				\mc D_{\underline{\alpha},\alpha_0}(\M,\Lambda,\vec{O}_X)=
				\log \mathcal{W}_{\mathbb{UI}}(\M,\Lambda,\vec{O}_X,u_{\vec R}).
			\end{equation}
		\end{result}
	\end{widetext}
	
	The proof is presented in \cref{a:r4}.
	This result proposes a resource measure $\mc D_{\underline{\alpha},\alpha_0}$ for measurement informativeness based on the difference between the unconditional and conditional multivariate R\'enyi divergences.
	The measure is also economic-theoretically meaningful; in terms of \cref{eq:ratio3}, it reflects the natural intuition that using informative measurements is more beneficial to the gambler in a multi-lottery SB game than using uninformative measurements.
	In other words, the result can be understood as providing an operational, physical interpretation of the data processing inequality for multivariate R\'enyi divergences, which was initially seen as an information-theoretic relation, through multi-lottery SB games.
	It is again notable that the data processing inequality (\cref{c:c1}) for multivariate R\'enyi divergences plays a crucial role when showing the monotonicity \cref{eq:RT_monotone}. 
	Now the order conditions \cref{eq:parameter} together with an additional requirement  $\alpha_0=\max_{k}\alpha_k$ are not only rephrased as the economic-theoretically consistent behaviour of a risk-averse gambler towards all lotteries but also lead to a new quantifier of how beneficial a measurement is in any GPT, the most general framework of physical theory.
	We also note that the measure $\mc D_{\underline{\alpha},\alpha_0}$ reproduces a \emph{complete set of monotones} for measurement informativeness in certain cases.
	This can be confirmed by recalling that the set of success probabilities $\{p^{\rm SD}_{\rm succ}(\M,\Lambda)\mid\Lambda\}$ of SDs in a GPT is a complete set of monotones; 
	for two measurements $\M_1$ and $\M_2$, there exists a postprocessing map $T\in \mc O_{\mathbb{UI}}$ such that $\M_2=T(\M_1)$ if and only if $p^{\rm SD}_{\rm succ}(\M_1,\Lambda)\ge p^{\rm SD}_{\rm succ}(\M_2,\Lambda)$ for every state ensemble $\Lambda$ \cite{PS_NL_2019, RT_BS_2019}.
	Then, by setting $\underline{\alpha}=(1,0,0\ldots,0)$ and applying constant odds, we can reduce the ICE $w^{\rm ICE}_{R=0}(\M,\Lambda,s_{G|A},C,b^{(1)}_{X|G})$ of the corresponding SB to the success probability $p^{\rm SD}_{\rm succ}
	(\M,\Lambda)$  (see \cref{eq:ICE-SD}) and thus verify the claim.
	
	\section{Summary and outlook}
	\label{s:conclusions}
	In this work we showed that multivariate R\'enyi divergences characterise multi-lottery betting games. 
	Specifically, we revealed that the unconditional multivariate R\'enyi divergence represents the economic-theoretic value that a rational agent (gambler) assigns to an operational task based on risk aversion and betting on multiple lotteries when the odds are fair, and when the gambler is allowed to maximise over all betting strategies. 
	Furthermore, we studied a generalised scenario where the gambler is allowed to implement betting strategies after having access to side information. 
	The analysis of this extended scenario led us to introduce a new \emph{conditional} multivariate R\'enyi divergence, as the appropriate quantity that describes this generalised scenario involving side information. 
	In particular, we derived a data processing inequality for this new conditional multivariate R\'enyi divergence, and proved that it can be endowed with an economic-theoretic interpretation; having access to side information increases the economic-theoretic value that the gambler assigns to the lotteries in question. 
	This latter statement is somewhat intuitive, and can now be grounded in a clean mathematical statement of the data processing inequality. 
	It is also particularly notable that the order restrictions for the conditional multivariate R\'enyi divergence to satisfy the data processing inequality are derived from the gambler's economic-theoretic consistency of risk aversion for all lotteries.
	More precisely, we proved the quantitative equivalency between the gambler's treatment of every lottery with a similar sensitivity to risk aversion and the mathematical requirements for the order parameters of the information-theoretically proper divergence.
	Finally, we incorporated these results into the scenario of multi-lottery state betting games and studied their implications for the resource theory of informative measurements within the operational framework of general probabilistic theories (GPTs). 
	It was revealed that multivariate R\'enyi divergences define a resource monotone of measurement informativeness and that the monotone can be interpreted as expressing the economic-theoretic benefit of an informative measurement for a gambler when playing a multi-lottery state betting game.
	These observations were obtained in the realm of GPTs, which are physical theories with minimal operational requirements.
	Consequently, they establish a quantitative connection between information, economic, and physical theories with a high degree of generality.
	
	An interesting future direction of this study is to develop physical or other practical aspects of the majorisation property of the multivariate R\'enyi divergence.
	In \cite{farooq2024}, a curious characterisation of the unconditional multivariate R\'enyi divergence was given stating that a set $\vec{P}_X$ of PMFs asymptotically majorises another $ \vec{Q}_X$ (in our language, $\vec{Q}_X$ is obtained by postprocessing of $\vec{P}_X$) with large samples if and only if $D_{\underline{\alpha}}(\vec{P}_X)\ge D_{\underline{\alpha}}(\vec{Q}_X)$ holds for all the admissible orders $\underline{\alpha}$.
	It is natural to expect this majorisation property to be possibly incorporated into the framework of this study.
	In particular, to investigate whether the conditional multivariate R\'enyi divergence exhibits a similar property is not only an information-theoretically motivated problem but may also yield several economic-theoretic insights in a quantitative way.
	From the resource-theoretic perspective, the problem may be related to convertibility of resource, which is a central problem in the field of resource theory.
	It is also an interesting future problem to investigate whether a set of monotones $\{\mc D_{\underline{\alpha},\alpha_0}(\M,\Lambda,\vec{O}_X)\mid \Lambda, \vec{O}_X\}$ introduced in \cref{r:r4} is a complete set of monotones for a fixed $\underline{\alpha}$.
	This set generalises the complete set of monotones given by state discriminations, and this problem will bring another operational and practical interpretation to the field of resource theory.

    Our resource-theoretic results involve entropic quantities obtained from measurement outcome statistics. This is why these quantities are somewhat ``classical'' in nature. When studying the resource theory of (ensembles of) quantum states or quantum channels (completely positive and trace-preserving linear maps on operators), it seems natural that the relevant resource monotones are ``more quantum'' in nature. Such quantities could be related to quantum relative entropies \cite{Petz_85,Petz_1986,Hiai_Petz_91,Muller-Lennert_et_al_2013,Jaksic2012,Audenaert_Datta_2015} or their multivariate extensions \cite{Furuya_et_al_2023,mosonyi2024geometric,Nuradha2025}; see also Section 4 of \cite{MV4} and references therein. However, the study of the latter is still in its infancy and investigation into {\it conditional} multivariate quantum divergences is, to the best of our knowledge, completely non-existent. Some quantum divergences, like the ``sandwiched'' quantum relative entropies \cite{Muller-Lennert_et_al_2013}, can be obtained as optimised and regularised measured (thus ``classical'') quantities, see e.g.\ Proposition 4.12 of \cite{Tomamichel2016Book}. It remains to be seen if similar techniques might work in the multivariate (and conditional) settings, yielding meaningful resource monotones.
	
	\vspace{7mm}
	
	\section*{Acknowledgements}
	A.F.D. thanks Josep Lumbreras, Ruo Cheng Huang, Jeongrak Son, Xiangjing Liu, Jayne Thompson, and Mile Gu, for interesting discussions on betting games.
	A.F.D. acknowledges support from the National Research Foundation of Singapore through the NRF Investigatorship Program (Award No. NRF-NRFI09-0010), the National Quantum Office, hosted in A*STAR, under its Centre for Quantum Technologies Funding Initiative (S24Q2d0009), the International Research Unit of Quantum Information, Kyoto University, the Center for Gravitational Physics and Quantum Information (CGPQI), and COLCIENCIAS 756-2016.
    E.H.\ acknowledges support from the National Research Foundation Investigatorship Award (NRF-NRFI10-2024-0006) and from the National Research Foundation, Singapore through the National Quantum Office, hosted in A*STAR, under its Centre for Quantum Technologies Funding Initiative (S24Q2d0009). 
	R.T. acknowledges support from MEXT QLEAP, JST COI-NEXT program Grant No. JPMJPF2014, and JSPS KAKENHI Grant No. JP25K17314.

	
	\onecolumngrid
	\appendix
	\hrulefill
	\crefalias{section}{appendix}
	\crefalias{subsection}{appendix}
	\section{Multivariate R\'enyi divergences for continuous sample spaces}
	\label{a:contrem}
	
	Let us make a couple of discussions regarding general sample spaces. 
	We first study the unconditional multivariate R\'{e}nyi divergence in \cref{def:divergences} on a measurable space $(\mc X,\Sigma)$, where $\Sigma$ is a $\sigma$-algebra of subsets of an alphabet $\mc X$.
	We consider probability measures $p^{(k)}_X\colon\Sigma\to [0,1]$ $(k=0,\ldots,d)$ on $(\mc X,\Sigma)$.
	When treating probability densities on a general sample space, we need a reference measure $M$ (e.g. the Lebesgue measure) on it.
	Here we set $M=p^{(0)}_X$ for simplicity, but generalising the following observations to a general $M$ is straightforward by formally replacing $dp^{(0)}_X$ with $dM$.
	Now we assume that all other probability measures $p$ are dominated by $p^{(0)}_X$, i.e., whenever $p^{(0)}_X(A)=0$ for some $A\in\Sigma$ then $p(A)=0$.  
	This means that $p$ is absolutely continuous with respect to $p^{(0)}_X$ (expressed as $p\ll p^{(0)}_X$), so that we may define the Radon-Nikod\'{y}m derivative (density) $dp/dp^{(0)}_X$ of $p$ with respect to $p^{(0)}_X$. 
	We still denote the expectation value of a measurable function $f_X:\mc X\to\R$ with respect to $p^{(0)}_X$ by
	$
	\mb E_{p^{(0)}_X}[f_X]\coloneqq \int_X dp^{(0)}_X(x) f_X(x)
	$.
	In both cases of \cref{eq:parameter}, 
	we may define the multivariate R\'{e}nyi divergence with orders  $\underline{\alpha}=(\alpha_0,\ldots,\alpha_{d})$
	for a family of probability measures (a \emph{statistical experiment} \cite{Torgersen1991}) $\vec P_X=\big(p^{(0)}_X,\ldots,p^{(d)}_X\big)$ with $p^{(k)}_X\ll p^{(0)}_X$ ($k=1,\ldots,d$) through
	\begin{equation*}
		D_{\underline{\alpha}}
		(\vec P_X)
		=
		\frac{1}{\alpha_{\star}-1}
		\log{
			\mb 
			E_{p^{(0)}_X}
			\left[
			\prod_{k=1}^{d}\left(\frac{dp^{(k)}_X}{dp^{(0)}_X}\right)^{\alpha_k}
			\right]
		},
	\end{equation*}
	where $\alpha_{\star}:=\max_{0\leq k\leq d}\alpha_k$.
	Moreover, the (bivariate) R\'{e}nyi divergence of degree $S\in[0,1)\cup(1,\infty)$ for $p_X^{(1)},p_X^{(2)}\ll p^{(0)}_X$ is defined through
	\begin{equation*}
		D_S(p_X^{(1)}\|p_X^{(2)})=\frac{1}{S-1}\log{\mb E_{p^{(0)}_X}\left[\left(\frac{dp_X^{(1)}}{dp^{(0)}_X}\right)^{S}\left(\frac{dp_X^{(2)}}{dp^{(0)}_X}\right)^{1-S}\right]}.
	\end{equation*}
	This divergence has to be set to infinity when the supports of $p_X^{(1)}$ and $p_X^{(0)}$ do not intersect and $S\in[0,1)$, and when $p_X^{(1)}$ has support outside that of $p_X^{(2)}$ and $S>1$. Regarding the support conditions, however, we essentially need to assume that $(\mc X,\Sigma)$ is standard Borel.

	Similar arguments can be applied to the conditional case developed in Definition \ref{def: cond mv Renyi}.
	We assume that the sample spaces are measurable spaces $(\mc X,\Sigma)$ and $(\mc G,\Gamma)$ where the latter describes side information. We will typically assume that these measurable spaces are standard Borel to make things simpler. The total sample space is $(\mc X\times\mc G,\Sigma\otimes\Gamma)$ where $\Sigma\otimes\Gamma$ is the coarsest $\sigma$-algebra on $\mc X\times\mc G$ which contains the product sets $A\times K$ where $A\in\Sigma$ and $K\in\Gamma$.
	We fix a probability measure $p^{(0)}_{}:\Sigma\otimes\Gamma\to[0,1]$ which dominates all other measures on $G$ that we consider.
	We define the additional probability measures
	$p^{(0)}_X:\Sigma\to[0,1]$ and $p^{(0)}_G:\Gamma\to[0,1]$ through
	$$
	p^{(0)}_X(A)=p^{(0)}(A\times \mc G),\quad p^{(0)}_G(K)=p^{(0)}(\mc X\times K)
	$$
	for all $A\in\Sigma$ and $K\in\Gamma$.
	We may also define the positive measures $p^{(0)}_{A,G}:\,K\mapsto p^{(0)}(A\times K)$ for all $A\in\Sigma$. These measures are all absolutely continuous with respect to $p^{(0)}_G$ as one easily sees. 
	It enables us to define the function $p^{(0)}_{X|G}(\cdot|\cdot):\Sigma\times \mc G\to[0,\infty]$ through
	$$
	p^{(0)}_{X|G}(A|g)=\frac{dp^{(0)}_{A,G}}{dp^{(0)}_G}(g),\quad A\in\Sigma,\ g\in \mc G.
	$$
	It is easy to see that $p^{(0)}_{X|G}(\cdot|g)$ is a probability measure for all $g\in \mc G$, i.e.,\ $p^{(0)}_{X|G}(\cdot|\cdot)$ is a Markov kernel. It follows that
	$$
	p^{(0)}(A\times K)=\int_K p^{(0)}_{X|G}(A|g)\,dp^{(0)}_G(g),
	\quad 
	A\in\Sigma,\ K\in\Gamma.
	$$
	Let us pick $\alpha\geq0$ in addition to the parameter vector $\underline{\alpha}=(\alpha_0,\ldots,\alpha_{d})$. 
	For any probability measures $p^{(k)}_{X|G}:\Sigma\to[0,1]$ ($k=1,\ldots,d$) such that $p^{(k)}_{X|G}(\cdot|g)\ll p^{(0)}_{X|G}(\cdot|g)$ for all $g\in \mc G$, we can define the continuous extension of the conditional multivariate R\'{e}nyi divergence (with respect to $p^{(0)}_G$) through
	\begin{align*}
		D_{\underline{\alpha},\alpha}
		\big(
		p^{(0)}_{X|G},\ldots,p^{(d)}_{X|G}\big|p^{(0)}_{G}
		\big)
		=
		\frac{\alpha}{\alpha_\star-1}\log\Bigg\{\!\int_{\mc G}dp^{(0)}_G(g)
		\Bigg(\!
		\int_{\mc X}p^{(0)}_{X|G}(dx|g)\prod_{k=1}^{d}\bigg(\frac{dp^{(k)}_{X|G}(\cdot|g)}{dp^{(0)}_{X|G}(\cdot|g)}(x)\bigg)^{\!\alpha_k}
		\Bigg)^{\!1/\alpha}\Bigg\},
	\end{align*}
	where $\alpha_{\star}=\max_{0\leq k\leq d}\alpha_k$.

	\section{Monotonicity of multivariate R\'enyi divergences for order vectors}
	\label{a:l2-2}
\begin{figure}[h!]
		\begin{center}
			\vspace{3mm}
			\begin{overpic}[scale=0.4, unit=1mm]{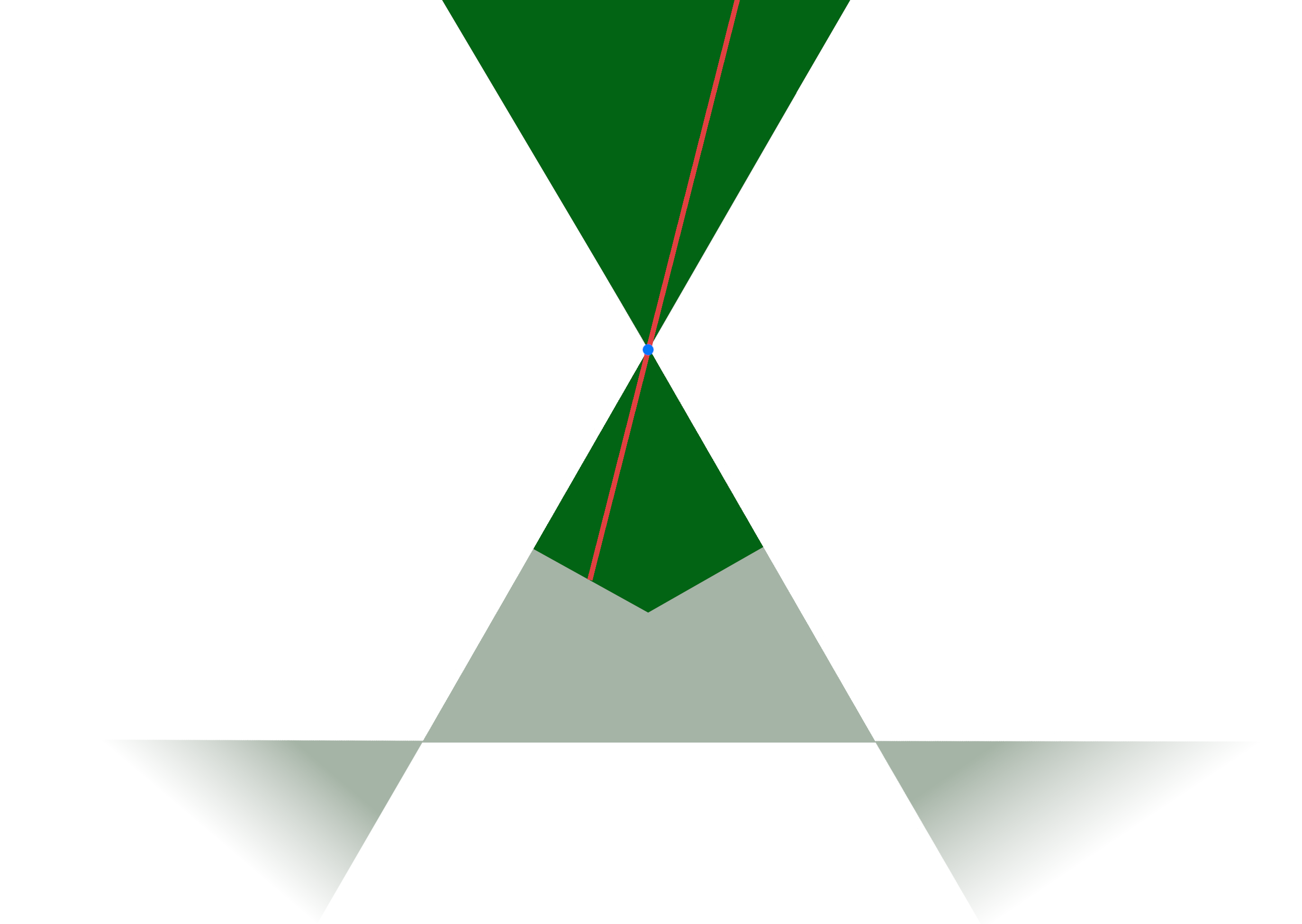}
				\put(39,40){\begin{large}
						$e_0$
				\end{large}}
				\put(24,14){\begin{large}
						$e_1$
				\end{large}}
				\put(62,14){\begin{large}
						$e_2$
				\end{large}}
				\put(50,68){\begin{large}
						\punainen{$\underline{\alpha}^\lambda$}
				\end{large}}
			\end{overpic}
		\end{center}
		\caption{\label{fig:parameterregion_app}The area of allowed parameter vectors $\underline{\alpha}=(\alpha_0,\alpha_1,\alpha_2)\in\R^3$ in the case $d=2$. The area is confined onto the affine plane $\sum_{k=0}^2\alpha_k=1$. The special points fixing this plane are $e_0=(1,0,0)$, $e_1=(0,1,0)$, and $e_2=(0,0,1)$. The total shaded area represents all \cref{eq:parameter} and the darker area is where $\alpha_0$ is the largest among the $\alpha_k$. The path of \cref{eq:path} satisfying \cref{eq:monotone} is depicted in red with the special point $\lambda=1$ at $e_0$ in blue. As we move upward along this path, the value $D_{\underline{\alpha}^\lambda}(P)$ increases asymptotically approaching the limit quantity $D^\T_{\underline{\beta}}(P)$ where $\underline{\beta}$ depends on the angle of the path.
		}
	\end{figure}
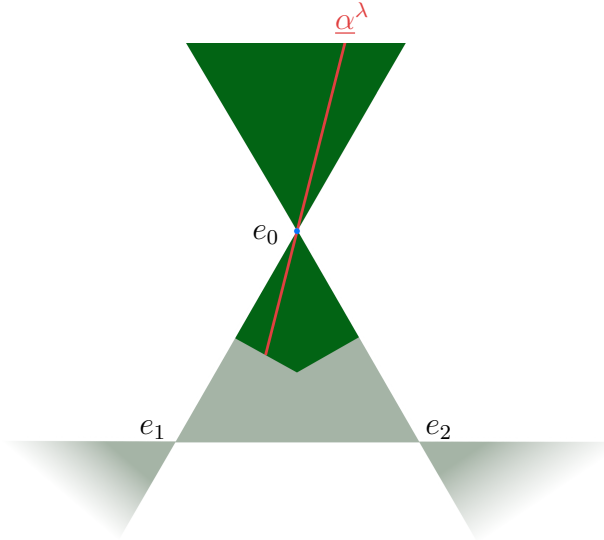

	We may also identify certain monotonicity properties for the unconditional multivariate R\'enyi divergence $D_{\underline{\alpha}}(\vec{P}_X)$ as a function of $\underline{\alpha}$ in the allowed region for any fixed $\vec P_X=\big(p^{(0)}_X,\ldots,p^{(d)}_X\big)$ \cite[Proposition 17]{farooq2024}.
	Let us remain in the region of \cref{eq:parameter} with $\alpha_0$ the largest among $k=0,\ldots,d$; the other sections of the parameter region can be dealt with in a similar fashion. Let us fix $\gamma_1,\ldots,\gamma_{d}\geq0$ such that $\gamma_1+\cdots+\gamma_{d}=1$ and consider the path
	\begin{equation}\label{eq:path}
		\begin{aligned}
			&[0,1)\cup(1,\infty)\ni\lambda
			\mapsto
			\underline{\alpha}^\lambda=\big(\lambda,(1-\lambda)\gamma_1,\ldots,(1-\lambda)\gamma_{d}\big)\in\R^{d+1}
		\end{aligned}
	\end{equation}
	within the parameter space.
	We illustrate the setting for the case $d=3$ in \cref{fig:parameterregion_app}.
	Note that we set 
	\begin{equation}\label{eq:monotone}
		\lambda\geq\max_{1\leq k\leq d}\frac{\gamma_k}{\gamma_k+1},\quad \lambda\neq1,
	\end{equation}
	so that $\alpha_0=\lambda$ is the largest among the components of $\underline{\alpha}^\lambda$ and they are within the region characterised by \cref{eq:parameter}. 
	In the region described by \cref{eq:monotone}, the divergence $D_{\underline{\alpha}^\lambda}(\vec P_X)$ is non-decreasing in $\lambda$ for any fixed $\vec P_X$. The pointwise limit at $\lambda\to 1$ exists and is given by
	$$
	\lim_{\lambda\to 1}D_{\underline{\alpha}^\lambda}(\vec P_X)=\sum_{k=1}^{d}\gamma_k D_{\rm KL}\big(p^{(0)}_X\big\|p^{(k)}_X\big),
	$$
	where $D_{\rm KL}$ is the Kullback-Leibler divergence
	$$
	D_{\rm KL}(p_X\|q_X)=\mb E_{q_X}\left[\frac{p_X}{q_X}\log{\frac{p_X}{q_X}}\right].
	$$
	Also the limit $\lambda\to\infty$ exists and is given by
	\begin{align*}
		\lim_{\lambda\to\infty}D_{\underline{\alpha}^\lambda}(\vec P_X)
		=\sup_{\lambda>1}D_{\underline{\alpha}^\lambda}(\vec P_X)
		&=D^\T_{(1,-\gamma_1,\ldots,-\gamma_{d})}(\vec P_X)\\
		&:=\max_{x\in \mc X}~\log{\frac{p^{(0)}_X(x)}{(p^{(1)}_X(x))^{\gamma_1}\cdots (p^{(d)}_X(x))^{\gamma_{d}}}}.
	\end{align*}
	The limit quantities $D^\T_{\underline{\beta}}$ for any $\underline{\beta}=(\beta_0,\ldots,\beta_{d})\in\R^{d+1}$ with $\beta_0>0$, $\beta_1,\ldots,\beta_{d}\leq0$, and $\beta_0+\cdots+\beta_{d}=0$ satisfy the data processing inequality as well as many other desirable properties \cite{farooq2024}.

	\section{Proof of \cref{l:l2}} 
	\label{a:l2}
	\begin{proof}
		Let us first assume that condition (i) of \cref{eq:parameter} is satisfied. 
		Using Minkowski's inequality, we now have
		\begin{align*}
			\sum_{h\in \mc H}q^{}_H(h)&\left(\sum_{x\in\mc X}\left(q^{(0)}_{X|H}(x|h)\prod_{k=1}^{d} \big(r^{(k)}_X(x)\big)^{\alpha_k/\alpha_0}\right)^{\alpha_0}\right)^{1/\alpha_0}
			\\
			&\qquad\qquad=\sum_{h\in\mc H}q^{}_H(h)\left(\sum_{x\in\mc X}\left(\sum_{g\in\mc G}t^\dagger_{G|H}(g|h)p^{(0)}_{X|G}(x|g)\prod_{k=1}^{d} \big(r^{(k)}_X(x)\big)^{\alpha_k/\alpha_0}\right)^{\alpha_0}\right)^{1/\alpha_0}\\
			&\qquad\qquad\leq\sum_{g\in\mc G}\sum_{h\in\mc H}\underbrace{q^{}_{H}(h)t^\dagger_{G|H}(g|h)}_{=p^{}_G(g)t_{H|G}(h|g)}\left(\sum_{x\in\mc X}\left(p^{(0)}_{X|G}(x|g)\prod_{k=1}^{d} \big(r^{(k)}_X(x)\big)^{\alpha_k/\alpha_0}\right)^{\alpha_0}\right)^{1/\alpha_0}\\
			&\qquad\qquad=\sum_{g\in\mc G}p^{}_G(g)\underbrace{\sum_{h\in\mc H}t_{H|G}(h|g)}_{=1}\left(\sum_{x\in\mc X}\left(p^{(0)}_{X|G}(x|g)\prod_{k=1}^{d} \big(r^{(k)}_X(x)\big)^{\alpha_k/\alpha_0}\right)^{\alpha_0}\right)^{1/\alpha_0}\\
			&\qquad\qquad=\sum_{g\in\mc G}p^{}_G(g)\left(\sum_{x\in\mc X}\left(p^{(0)}_{X|G}(x|g)\prod_{k=1}^{d} \big(r^{(k)}_X(x)\big)^{\alpha_k/\alpha_0}\right)^{\alpha_0}\right)^{1/\alpha_0}.
		\end{align*}
		In the case (ii), the inequality on the third line above is reversed. Noting that $\alpha_0/(\alpha_0-1)$ is negative in case (i) and positive in case (ii), we now have, in both cases,
		\begin{align*}
			D_{\underline{\alpha},\alpha_0}\big(q^{(0)}_{X|H},r_X^{(1)},\ldots,r^{(d)}_{X}\big|q^{}_H\big)
			&=\frac{\alpha_0}{\alpha_0-1}\log\sum_{h\in \mc H}q^{}_H(h)\left(\sum_{x\in\mc X}\left(q^{(0)}_{X|H}(x|h)\prod_{k=1}^{d} \big(r^{(k)}_X(x)\big)^{\alpha_k/\alpha_0}\right)^{\alpha_0}\right)^{1/\alpha_0}\\
			&\leq\frac{\alpha_0}{\alpha_0-1}\log{\sum_{g\in\mc G}p^{}_G(g)\left(\sum_{x\in\mc X}\left(p^{(0)}_{X|G}(x|g)\prod_{k=1}^{d} \big(r^{(k)}_X(x)\big)^{\alpha_k/\alpha_0}\right)^{\alpha_0}\right)^{1/\alpha_0}}\\
			&=D_{\underline{\alpha},\alpha_0}\big(p^{(0)}_{X|G},r_X^{(1)},\ldots,r^{(d)}_{X}\big|p^{}_G\big),
		\end{align*}
		proving \cref{eq:BLPineq}.
	\end{proof}
	\section{Proofs of main results}
	\subsection{Proof of \cref{r:r1}}
	\label{a:r1}
	\begin{proof}
		We simply write $w^{\rm ICE}
		(p_X^{(0)},\vec{O}_X,\vec{B}_X, u_{\underline{R}})$ by $w^{\rm ICE}$.
		Since $\sum_{k=1}^{d}(1-R_k)=(\alpha_0-1)/\alpha_0$,  
		solving from \cref{eq:multiCE} under \cref{eq:multi-IEU} implies
		\begin{align}
			\log{w^{\rm ICE}}
			=\frac{\alpha_0}{\alpha_0-1}\log\sum_{x\in \mc X}\big(b^{(1)}_X(x)\big)^{1-S_1}\left(C_1\cdot q^{(1)}_X(x)\right)^{S_1},\label{eq:1}
		\end{align}
		where we introduced numbers $S_1:=(\alpha_0+\alpha_1)/\alpha_0\geq0$,
		\begin{align*}
			C_1:=
			\sum_{x\in \mc X}\left\{\big(p_{X}^{(0)}(x)\big)^{\alpha_0}
			\left(\frac{1}{o^{(1)}_X(x)}\right)^{\alpha_1}\prod_{k=2}^{d}\big(b^{(k)}_X(x)o^{(k)}_X(x)\big)^{-\alpha_{k}}
			\right\}^{\frac{1}{\alpha_0+\alpha_1}}
		\end{align*}
		and a function
		\begin{align*}
			q^{(1)}_X
			:=\frac{1}{C_1}\left\{\big(p_{X}^{(0)}\big)^{\alpha_0}
			\left(\frac{1}{o^{(1)}_X}\right)^{\alpha_1}\prod_{k=2}^{d}\big(b^{(k)}_Xo^{(k)}_X\big)^{-\alpha_{k}}
			\right\}^{\frac{1}{\alpha_0+\alpha_1}}
		\end{align*}
		on $\mc X$.
		Note that the constant $C_1$ is defined so that $q^{(1)}_X$ is a PMF and also that $q^{(1)}_X$ is independent of $b^{(1)}_X$.
		Thus we can rephrase \cref{eq:1} by means of the bivariate R\'enyi divergence \cref{eq:Renyi_bi} as 
		\begin{align}
			\log{w^{\rm ICE}}
			&\!=\!\frac{\alpha_1}{\alpha_0-1}D_{S_1}\big(q_X^{(1)}\big\|b_X^{(1)}\big)+\frac{\alpha_0+\alpha_1}{\alpha_0-1}\log{C_1}.\label{eq:peal1}
		\end{align}
		We apply the same argument to the term $\log{C_1}$ to peal off the PMF $b_X^{(2)}$ and obtain
		\begin{align*}
			\log{C_1}
			&=\log\sum_{x\in \mc X}\big(b^{(2)}_X(x)\big)^{1-S_2}\left(C_2\cdot q^{(2)}_X(x)\right)^{S_2}\\
			&=\frac{\alpha_2}{\alpha_0+\alpha_1}D_{S_2}\big(q_X^{(2)}\big\|b_X^{(2)}\big)+\frac{\alpha_0+\alpha_1+\alpha_2}{\alpha_0+\alpha_1}\log{C_2},
		\end{align*}
		where $S_2:=(\alpha_0+\alpha_1+\alpha_2)/(\alpha_0+\alpha_1)\geq0$,
		$$
		C_2:=\sum_{x\in \mc X}\left\{\big(p_{X}^{(0)}(x)\big)^{\alpha_0}\left(\frac{1}{o^{(1)}_X(x)}\!\right)^{\alpha_1}\left(\frac{1}{o^{(2)}_X(x)}\right)^{\alpha_2}\prod_{k=3}^{d}\big(b_X^{(k)}o_X^{(k)}\big)^{-\alpha_{k}}\right\}^{\frac{1}{\alpha_0+\alpha_1+\alpha_2}}
		$$
		are constants and
		$$
		q_X^{(2)}:=\frac{1}{C_2}
		\sum_{x\in\mc X}\left\{\big(p_{X}^{(0)}\big)^{\alpha_0}
		\left(\frac{1}{o^{(1)}_X}\right)^{\alpha_1}\left(\frac{1}{o^{(2)}_X}\right)^{\alpha_2}\prod_{k=2}^{d}\big(b^{(k)}_Xo^{(k)}_X\big)^{-\alpha_{k}}
		\right\}^{\frac{1}{\alpha_0+\alpha_1+\alpha_2}}
		$$
		is a PMF.
		Applying the same argument and pealing off the bet functions $b^{(k)}_X$ one by one, 
		we can reduce \cref{eq:peal1} to 
		\begin{align}
			\log{w^{\rm ICE}}=\frac{1}{\alpha_0-1}\left(\sum_{k=1}^{d}\alpha_k D_{S_{k}}\big(q_X^{(k)}\big\|b_X^{(k)}\big)+\log{C_{d}}\right),\label{eq:apu0}
		\end{align}
		with constants
		\begin{align*}
			&S_k:=\frac{\alpha_0+\cdots+\alpha_{k}}{\alpha_0+\cdots+\alpha_{k-1}},\\
			&C_{k}:=
			\sum_{x\in \mc X}\left\{\big(p_{X}^{(0)}(x)\big)^{\alpha_0}\prod_{l=1}^{k}\left(\frac{1}{o^{(l)}_X(x)}\right)^{\alpha_l}\prod_{m=k+1}^{d}\big(b_X^{(m)}(x)o_X^{(m)}(x)\big)^{-\alpha_{m}}\right\}^{\frac{1}{\alpha_0+\cdots+\alpha_k}},
		\end{align*} 
		and PMFs 
		\begin{align*}
			q_X^{(k)}:=
			\frac{1}{C_k}\left\{\big(p_{X}^{(0)}\big)^{\alpha_0}\prod_{l=1}^{k}\left(\frac{1}{o^{(l)}_X}\right)^{\alpha_l}\prod_{m=k+1}^{d}\big(b_X^{(m)}o_X^{(m)}\big)^{-\alpha_{m}}\right\}^{\frac{1}{\alpha_0+\cdots+\alpha_k}}
		\end{align*} 
		for $k=1,\ldots,d$.
		In particular, 
		\begin{align*}
			\log{C_{d}}
			&=\log\sum_{x\in \mc X}\big(p_{X}^{(0)}(x)\big)^{\alpha_0}\prod_{k=1}^{d}\left(\frac{1}{o_X^{(k)}(x)}\right)^{\alpha_k}\\			
			&=\sum_{k=1}^{d}\alpha_k\log{F^{(k)}}
			+(\alpha_0-1)D_{\underline{\alpha}}(p_X^{(0)},p_X^{(1)},\ldots,p_X^{(d)})
		\end{align*}
		holds, and substituting this in \cref{eq:apu0}, we obtain \cref{eq:wCEform}.
	\end{proof}
	
	The continuous counterpart of the $d$-commodity betting game can be described in the same way as \cref{s:s4a}.		
	Now each bet $b^{(k)}_X$ is a probability measure and odds function $o_X^{(k)}$ is a measurable function on $X$.
	Moreover, we assume that there are probability measures $p^{(k)}_X$ and $F^{(k)}>0$, $k=1,\ldots,d$, such that $(o^{(k)}_X)^{-1}=F^{(k)}dp^{(k)}_X/dp^{(0)}_X$. 
	Setting again $p_X^{(0)}$ as the dominating measure, we can develop the same proof as \cref{r:r1} with formally replacing $\sum_x\to\int dp^{(0)}_X$ and $b_X^{(k)}\to db^{(k)}_X/dp^{(0)}_X$ (also $(p_{X}^{(0)}(x))^{\alpha_0}\to (dp^{(0)}_X/dp^{(0)}_X)^{\alpha_0}=1$).
	It can be finally proved that there are probability measures $q^{(1)}_X,\ldots,q^{(d)}_X\ll p^{(0)}_X$ and $S_1,\ldots,S_{d}\geq0$ such that we have for the isoelastic certainty equivalent $w^{\rm ICE}$ in the utility function \cref{eq:multi-IEU} and the statistical experiment $\vec{P}_X:=\big(p_X^{(0)},\ldots,p_X^{(d)}\big)$
	\begin{align*}
		\log{w^{\rm ICE}}
		&=D_{\underline{\alpha}}(\vec{P}_X)+\frac{1}{\alpha_0-1}\sum_{k=1}^{d} \alpha_k\left\{D_{S_{k}}\big(q^{(k)}_X\big\|b^{(k)}_X\big)+\log{F^{(k)}}\right\}\\
		&\leq D_{\underline{\alpha}}(\vec{P}_X)+\frac{1}{\alpha_0-1}\sum_{k=1}^{d} \alpha_k\log{F^{(k)}},
	\end{align*}
	where the upper bound is reached by optimising over the betting strategies.
	
	\subsection{Proof \cref{r:r2}} 
	\label{a:r2}
	\begin{proof}
		We write $w^{\rm ICE}
		(p^{(0)}_{XG}, \vec{O}_X, \vec{B}_{X|G},u_{\vec R})$ by $w^{\rm ICE}$.
		Proceeding as earlier, we have
		\begin{align}
			\log{w^{\rm ICE}}
			&=\frac{\alpha_0}{\alpha_0-1}\log{
				\sum_{(x,g)\in \mc X\times \mc G}p^{(0)}_{XG}(x,g)\left[\prod_{k=1}^{d}\big(b_{X|G}^{(k)}(x|g)o_X^{(k)}(x)\big)^{-\frac{\alpha_k}{\alpha_0}}\right]
			}\notag\\
			&=\frac{\alpha_0}{\alpha_0-1}\log\sum_{x,g}\big(b^{(1)}_{X|G}(x|g)\big)^{1-S_1}\left(c_{1}(g)\cdot q^{(1)}_{X|G}(x|g)\right)^{S_1},\label{eq:app_r2_1}
		\end{align}
		where we introduced numbers $S_1:=(\alpha_0+\alpha_1)/\alpha_0\geq0$,
		\begin{align*}
			c_{1}(g):=
			\sum_{x\in \mc X}\left\{\big(p_{XG}^{(0)}(x,g)\big)^{\alpha_0}
			\left(\frac{1}{o^{(1)}_X(x)}\right)^{\alpha_1}\prod_{k=2}^{d}\big(b^{(k)}_{X|G}(x|g)o^{(k)}_X(x)\big)^{-\alpha_{k}}
			\right\}^{\frac{1}{\alpha_0+\alpha_1}},
		\end{align*}
		a conditional PMF $q_{X|G}^{(1)}$ by
		\begin{align*}
			q_{X|G}^{(1)}(x|g):=\frac{1}{c_{1}(g)}
			\left\{\big(p_{XG}^{(0)}(x,g)\big)^{\alpha_0}
			\left(\frac{1}{o^{(1)}_X(x)}\right)^{\alpha_1}\prod_{k=2}^{d}\big(b^{(k)}_{X|G}(x|g)o^{(k)}_X(x)\big)^{-\alpha_{k}}
			\right\}^{\frac{1}{\alpha_0+\alpha_1}}
		\end{align*}
		that produces a PMF on $\mc X$ for each $g\in \mc G$.
		We in addition define a constant
		\begin{equation*}
			C_1=\sum_{g\in G} (c_{1}(g))^{S_1}=\sum_{g}\left[\sum_x\left\{\big(p_{XG}^{(0)}(x,g)\big)^{\alpha_0}
			\left(\frac{1}{o^{(1)}_X(x)}\right)^{\alpha_1}\prod_{k=2}^{d}\big(b^{(k)}_{X|G}(x|g)o^{(k)}_X(x)\big)^{-\alpha_{k}}
			\right\}^{\frac{1}{\alpha_0+\alpha_1}}\right]^{S_1},
		\end{equation*}
		and a PMF $q_G^{(1)}$ on $\mc G$ by
		\begin{equation*}
			q_G^{(1)}(g)=\frac{(c_{1}(g))^{S_1}}{C_1}
		\end{equation*}
		It enables us to rewrite \cref{eq:app_r2_1} as
		\begin{align}
			\log{w^{\rm ICE}}
			&=\frac{\alpha_0}{\alpha_0-1}\log \sum_{g}q^{(1)}_{G}(g)\sum_x\big(b^{(1)}_{X|G}(x|g)\big)^{1-S_1}\big(q^{(1)}_{X|G}(x|g)\big)^{S_1}+\frac{\alpha_0}{\alpha_0-1}\log{C_1}\notag\\
			&=\frac{\alpha_0}{\alpha_0-1}\log \sum_{g,x}\big(q^{(1)}_{G}(g)b^{(1)}_{X|G}(x|g)\big)^{1-S_1}\big(q^{(1)}_{G}(g)q^{(1)}_{X|G}(x|g)\big)^{S_1}+\frac{\alpha_0}{\alpha_0-1}\log{C_1}\notag\\
			&=\frac{\alpha_1}{\alpha_0-1}D_{S_1}\big(q_{XG}^{(1)}\big\|r_{XG}^{(1)}\big)+\frac{\alpha_0}{\alpha_0-1}\log{C_1},\label{eq:app_r2_3}
		\end{align}
		where $q_{XG}^{(1)}:=q_{G}^{(1)}q^{(1)}_{X|G}$ and $r_{XG}^{(1)}:=q^{(1)}_{G}(g)b^{(1)}_{X|G}$ are joint PMFs on $\mc X\times \mc G$ and
		$D_{S_1}\big(q_{XG}^{(1)}\big\|r_{XG}^{(1)}\big)$ is their bivariate R\'enyi divergence.
		We note that by choosing $b^{(1)}_{X|G}=q^{(1)}_{X|G}$, we can make the first term vanish.
		Let us analyse the final term in the above equation (see \cref{eq:app_r2_1}) when the parameters $\underline{\alpha}$ satisfy condition (i) of \cref{eq:parameter} as
		\begin{align*}
			\log{C_1}
			=&\log
			\sum_{g}\left[\sum_x\left\{\big(p_{XG}^{(0)}(x,g)\big)^{\alpha_0}
			\left(\frac{1}{o^{(1)}_X(x)}\right)^{\alpha_1}\prod_{k=2}^{d}\big(b^{(k)}_{X|G}(x|g)o^{(k)}_X(x)\big)^{-\alpha_{k}}
			\right\}^{\frac{1}{\alpha_0+\alpha_1}}\right]^{S_1}\\
			=&\log
			\sum_{g}p^{(0)}_G(g)\left[\sum_x\left\{\big(p^{(0)}_{X|G}(x|g)\big)^{\alpha_0}
			\left(\frac{1}{o^{(1)}_X(x)}\right)^{\alpha_1}\prod_{k=2}^{d}\big(b^{(k)}_{X|G}(x|g)o^{(k)}_X(x)\big)^{-\alpha_{k}}
			\right\}^{\frac{1}{\alpha_0+\alpha_1}}\right]^{S_1}\\
			=&\log
			\sum_{g}p^{(0)}_G(g)\left[\sum_x\big(b^{(2)}_{X|G}(x|g)\big)^{1-S_2}
			\Big(c_{2}(g)\cdot q^{(2)}_{X|G}(x|g)\Big)^{S_2}
			\right]^{S_1},
		\end{align*}
		where we introduced numbers $S_2:=(\alpha_0+\alpha_1+\alpha_2)/(\alpha_0+\alpha_1)\geq0$,
		\begin{align}
			\label{eq:app_r2_4}
			c_{2}(g):=
			\sum_{x\in \mc X}\left\{\big(p_{X|G}^{(0)}(x|g)\big)^{\alpha_0}
			\left(\frac{1}{o^{(1)}_X(x)}\right)^{\alpha_1}	\left(\frac{1}{o^{(2)}_X(x)}\right)^{\alpha_2}\prod_{k=3}^{d}\big(b^{(k)}_{X|G}(x|g)o^{(k)}_X(x)\big)^{-\alpha_{k}}
			\right\}^{\frac{1}{\alpha_0+\alpha_1+\alpha_2}},
		\end{align}
		a conditional PMF $q_{X|G}^{(2)}$ by
		\begin{align*}
			q_{X|G}^{(2)}(x|g):=\frac{1}{c_{2}(g)}
			\left\{\big(p_{X|G}^{(0)}(x|g)\big)^{\alpha_0}
			\left(\frac{1}{o^{(1)}_X(x)}\right)^{\alpha_1}	\left(\frac{1}{o^{(2)}_X(x)}\right)^{\alpha_2}\prod_{k=3}^{d}\big(b^{(k)}_{X|G}(\cdot|g)o^{(k)}_X(x)\big)^{-\alpha_{k}}
			\right\}^{\frac{1}{\alpha_0+\alpha_1+\alpha_2}}.
		\end{align*}
		With a constant 
		\begin{equation*}
			C_2=\sum_{g}(c_{2}(g))^{S_1S_2}p^{(0)}_G(g)
		\end{equation*}
		and a PMF
		\begin{equation}
			\label{eq:app_r2_7}
			q_G^{(2)}=\frac{(c_{2}(g))^{S_1S_2}p^{(0)}_G}{C_2},
		\end{equation}
		on $G$, it follows that
		\begin{align*}
			\log{C_1}
			&=
			\log\sum_{g}q^{(2)}_G(g)\left[\sum_x\big(b^{(2)}_{X|G}(x|g)\big)^{1-S_2}\big(q^{(2)}_{X|G}(x|g)\big)^{S_2}\right]^{S_1}+\log C_2\\
			&\ge
			\log\left[\sum_{g,x}q^{(2)}_G(g)\big(b^{(2)}_{X|G}(x|g)\big)^{1-S_2}\big(q^{(2)}_{X|G}(x|g)\big)^{S_2}\right]^{S_1}+\log C_2\\
			&=\log\left[\sum_{g,x}\big(q^{(2)}_G(g)b^{(2)}_{X|G}(x|g)\big)^{1-S_2}\big(q^{(2)}_G(g)q^{(2)}_{X|G}(x|g)\big)^{S_2}\right]^{S_1}+\log C_2\\
			&=S_1(S_2-1)D_{S_2}\big(q_{XG}^{(2)}\big\|r_{XG}^{(2)}\big)+\log{C_2},
		\end{align*}
		where $q_{XG}^{(2)}:=q_{G}^{(2)}q^{(2)}_{X|G}$ and $r_{XG}^{(2)}:=q^{(2)}_{G}(g)b^{(2)}_{X|G}$ are joint PMFs, and the inequality is due to Jensen's inequality and the fact that $S_1\geq 1$ in case (i) of \cref{eq:parameter}. 
		If we set $b^{(2)}_{X|G}=q^{(2)}_{X|G}$, we clearly reach the above lower bound.
		Since $S_1\leq 1$ in case (ii) of \cref{eq:parameter}, the inequality is reversed in this case. 
		Noting that $\alpha_0/(\alpha_0-1)<0$ in case (i) and $\alpha_0/(\alpha_0-1)\geq0$ in case (ii), we now have
		\begin{align*}
			\frac{\alpha_0}{\alpha_0-1}\log{C_1}
			&\le
			\frac{\alpha_0S_1}{\alpha_0-1}(S_2-1)D_{S_2}\big(q_{XG}^{(2)}\big\|r_{XG}^{(2)}\big)+\frac{\alpha_0}{\alpha_0-1}\log{C_2}\\
			&=\frac{\alpha_2}{\alpha_0-1}D_{S_2}\big(q_{XG}^{(2)}\big\|r_{XG}^{(2)}\big)+\frac{\alpha_0}{\alpha_0-1}\log{C_2}.
		\end{align*}
		and thus from \cref{eq:app_r2_3} 
		$$
		\log{w^{\rm ICE}}\leq
		\frac{\alpha_1}{\alpha_0-1}D_{S_1}\big(q_{XG}^{(1)}\big\|r_{XG}^{(1)}\big)+ \frac{\alpha_2}{\alpha_0-1}D_{S_2}\big(q_{XG}^{(2)}\big\|r_{XG}^{(2)}\big)+\frac{\alpha_0}{\alpha_0-1}\log{C_2}.
		$$
		Remember that when we choose $b^{(k)}_{X|G}$ so that $b^{(k)}_{X|G}=q^{(k)}_{X|G}$ for $k=1,2$, the above inequality is saturated and the two bivariate R\'{e}nyi divergences vanish.
		We proceed in the same way using Jensen's inequality and notice that
		\begin{equation}
			\label{eq:app_r2_8}
			\frac{\alpha_0}{\alpha_0-1}\log{C_{k-1}}\leq\frac{\alpha_{k}}{\alpha_0-1}D_{S_k}\big(q_{XG}^{(k)}\big\|r_{XG}^{(k)}\big)+\frac{\alpha_0}{\alpha_0-1}\log{C_k}
		\end{equation}
		for $k=2,\ldots,d$ in both cases (i) and (ii) of \cref{eq:parameter}, where the constants $S_k$ and $C_k$, and the PMFs $q_{XG}^{(k)}$ and $r_{XG}^{(k)}$ are defined as in the step above. 
		Explicitly, they are given as (see \cref{eq:app_r2_4,eq:app_r2_7})
		\begin{equation*}
			\begin{aligned}
				&S_k=\frac{\alpha_0+\cdots+\alpha_{k}}{\alpha_0+\cdots+\alpha_{k-1}}\geq0,\\
				&c_{k}(g):=
				\sum_{x\in \mc X}\left\{\big(p_{X|G}^{(0)}(x|g)\big)^{\alpha_0}
				\prod_{l=1}^{k}\left(\frac{1}{o^{(l)}_X(x)}\right)^{\alpha_l}	\prod_{m=k+1}^{d}\big(b^{(m)}_{X|G}(x|g)o^{(m)}_X(x)\big)^{-\alpha_{m}}\right\}^{\frac{1}{\alpha_0+\cdots+\alpha_k}},\\
				&C_k=\sum_{g\in\mc G}(c_{k}(g))^{S_1\cdots S_k}p^{(0)}_G(g)=\sum_{g\in\mc G}(c_{k}(g))^{\frac{\alpha_0+\cdots +\alpha_k}{\alpha_0}}p^{(0)}_G(g),\\
				&q_{X|G}^{(k)}(x|g):=\frac{1}{c_{k}(g)}
				\left\{\big(p_{X|G}^{(0)}(x|g)\big)^{\alpha_0}
				\prod_{l=1}^{k}\left(\frac{1}{o^{(l)}_X(x)}\right)^{\alpha_l}	\prod_{m=k+1}^{d}\big(b^{(m)}_{X|G}(x|g)o^{(m)}_X(x)\big)^{-\alpha_{m}}\right\}^{\frac{1}{\alpha_0+\cdots+\alpha_k}},\\
				&q_G^{(k)}(g)=\frac{(c_{k}(g))^{S_1\cdots S_k}p^{(0)}_G(g)}{C_k}=\frac{(c_{k}(g))^{\frac{\alpha_0+\cdots +\alpha_k}{\alpha_0}}p^{(0)}_G(g)}{C_k},\\
				&q_{XG}^{(k)}(x,g):=q_{G}^{(k)}(g)q^{(k)}_{X|G}(x|g),\qquad r_{XG}^{(k)}(x,g):=q^{(k)}_{G}(g)b^{(k)}_{X|G}(x|g).
			\end{aligned}
		\end{equation*}
		Again, if choosing $b^{(k)}_{X|G}=q^{(k)}_{X|G}$, the bivariate R\'{e}nyi divergence vanishes saturating the bound in \cref{eq:app_r2_8}. 
		This finally gives us
		\begin{align}
			\log{w^{\rm ICE}}&\leq \frac{1}{\alpha_0-1}\sum_{k=1}^{d} \alpha_k D_{S_{k}}\big(q^{(k)}_{XG}\big\|r_{XG}^{(k)}\big)+\frac{\alpha_0}{\alpha_0-1}\log{C_{d}}\label{eq:medial}\\
			&\leq\frac{\alpha_0}{\alpha_0-1}\log{C_{d}}\label{eq:final}
		\end{align}
		where the final upper bound can be reached.
		Let us investigate the final term $\log{C_{d}}$. 
		Recalling $p_X^{(k)}
		\coloneqq
		\frac{1}{F^{(k)}o_X^{(k)}
		}$ and $F^{(k)}:=\sum_{x}
		\frac{1}{o_X^{(k)}(x)}
		$, we now have
		\begin{align*}
			\frac{\alpha_0}{\alpha_0-1}\log{C_{d}}
			&=\frac{\alpha_0}{\alpha_0-1}\log{\sum_{g}p^{(0)}_G(g)(c_{d}(g))^{\frac{1}{\alpha_0}}}\\
			&=\frac{\alpha_0}{\alpha_0-1}\log{\sum_{g}p^{(0)}_G(g)\left(\sum_{x}\big(p_{X|G}^{(0)}(x|g)\big)^{\alpha_0}\prod_{k=1}^{d}\left(\frac{1}{o^{(k)}_X(x)}\right)^{\alpha_k}\right)^{1/\alpha_0}}\\
			&=\frac{\alpha_0}{\alpha_0-1}\log{\sum_{g}p^{(0)}_G(g)\left(\sum_{x}\big(p_{X|G}^{(0)}(x|g)\big)^{\alpha_0}\prod_{k=1}^{d}\big(F^{(k)}p_X^{(k)}(x)\big)^{\alpha_k}\right)^{1/\alpha_0}}\\
			&=\frac{1}{\alpha_0-1}\sum_{k=1}^{d}\alpha_k\log{F^{(k)}}
			+\frac{\alpha_0}{\alpha_0-1}\log{\sum_{g}p^{(0)}_G(g)\left(\sum_{x}\big(p_{X|G}^{(0)}(x|g)\big)^{\alpha_0}\prod_{k=1}^{d}\big(p_X^{(k)}(x)\big)^{\alpha_k}\right)^{1/\alpha_0}}\\
			&=\frac{1}{\alpha_0-1}\sum_{k=1}^{d}\alpha_k\log{F^{(k)}}
			+D_{\underline{\alpha},\alpha_0}\big(p_{X|G}^{(0)},p_{X}^{(1)},\ldots,p_{X}^{(d)}\big|p_G^{(0)}\big).
		\end{align*}
		Substituting this to \cref{eq:medial} and \cref{eq:final} gives \cref{r:r2}.
		We remark that the continuous counterpart of the current theorem can be obtained by the same procedure as explained in \cref{a:r1}.
	\end{proof}
	\subsection{Proof of \cref{r:r4}} 
	\label{a:r4}
	\begin{proof}
		To prove \cref{eq:ratio3}, choose an uninformative measurement $\NN=\{n_a\}_{a\in \mc B}$ where $n_b=\eta_B(b)u$ with the unit effect $u$ and a PMF $\eta_B$ on $\mc B$.
		We introduce a PMF
		$$
		q_{XB}^{(0)}[\NN,\Lambda](x,b)=p_X^{(0)}(x)\ang{n_b,\omega_x}=p_X^{(0)}(x)\eta_B(b),
		$$
		which particularly satisfies $q^{(0)}_{X|B}=q^{(0)}_X=p_X^{(0)}$.
		It is obvious that the conditional multivariate divergence $D_{\underline{\alpha},\alpha_0}  \big(q^{(0)}_{X|B}, p_X^{(1)},\ldots,p^{(d)}_{X}\big|q_B^{(0)}[\NN,\Lambda]\big)$ reduces to an unconditional multivariate divergence: 
		\begin{align*}
			D_{\underline{\alpha},\alpha_0}  \big(q^{(0)}_{X|B}, p_X^{(1)},\ldots,p^{(d)}_{X}\big|q_B^{(0)}[\NN,\Lambda]\big)
			&=\frac{\alpha_0}{\alpha_0-1}\log{\sum_{b}q_B(b)\left(\sum_{x}\big(q_{X|B}^{(0)}(x|b)\big)^{\alpha_0}\prod_{k=1}^{d} \big(p^{(k)}_X(x)\big)^{\alpha_k}\right)^{1/\alpha_0}}\\
			&=\frac{1}{\alpha_0-1}\log{\sum_{x}\prod_{k=0}^{d} \big(p^{(k)}_X(x)\big)^{\alpha_k}}=D_{\underline{\alpha}}  \big(p^{(0)}_{X}, p_X^{(1)},\ldots,p^{(d)}_{X}\big).
		\end{align*}
		Then \cref{r:r3} (or \cref{cor:r3}) implies 
		\begin{align*}
			&\log\ \frac{\displaystyle
				\max_{
					\vec{B}_{X|G},\,s_{G|A}
				}w^{\rm ICE}\big(\M,\Lambda,s_{G|A},\vec{O}_X,\vec{B}_{X|G},u_{\underline{ R}}\big)
			}
			{
				\displaystyle
				\max_{
					\vec{B}_{X|G},\,s_{G|B}
				}
				w^{\rm ICE}\big(\NN,\Lambda,s_{G|B},\vec{O}_X,\vec{B}_{X|G},u_{\underline{R}}\big)
			}\\
			&\qquad\qquad\qquad\qquad\qquad=
			D_{\underline{\alpha},\alpha_0}  \big(p^{(0)}_{X|A}, p_X^{(1)},\ldots,p^{(d)}_{X}\big|p_A^{(0)}[\M,\Lambda]\big)
			-
			D_{\underline{\alpha},\alpha_0}  \big(q^{(0)}_{X|B}, p_X^{(1)},\ldots,p^{(d)}_{X}\big|q_B^{(0)}[\NN,\Lambda]\big)\\
			&\qquad\qquad\qquad\qquad\qquad=
			D_{\underline{\alpha},\alpha_0}  \big(p^{(0)}_{X|A}, p_X^{(1)},\ldots,p^{(d)}_{X}\big|p_A^{(0)}[\M,\Lambda]\big)
			-
			D_{\underline{\alpha}}  \big(p^{(0)}_{X}, p_X^{(1)},\ldots,p^{(d)}_{X}\big).
		\end{align*}
		This holds independently of the uninformative measurement $\NN$ and thus proves \cref{eq:ratio3}.
		The resource-monotonic property is a direct implication of the data processing inequality \cref{l:l2}.
	\end{proof}
	
	\twocolumngrid
	\bibliography{bibliography_0119.bib}
	
\end{document}